\documentclass[fleqn,usenatbib]{mnras}

\usepackage{newtxtext,newtxmath}

\usepackage[T1]{fontenc}
\usepackage{ae,aecompl}
\usepackage{lastpage}

\usepackage{graphicx}	
\usepackage{amsmath}	

\pdfoutput=1
\usepackage{longtable,epsfig,verbatim,xspace,multirow,ulem,hhline}
\usepackage{float}
\usepackage[usenames,dvipsnames]{xcolor}
\usepackage{physics}
\usepackage{tikz}
\usetikzlibrary{arrows,babel}
\usepackage[labelfont={bf},labelsep=endash]{caption}
\usetikzlibrary{positioning,arrows}
\usetikzlibrary{decorations.pathmorphing}
\usetikzlibrary{decorations.markings}
\tikzset{
momentum/.style={postaction={decorate},decoration={markings,mark=at position 1 with {\arrow{>}}}},
particle/.style={dashed
    },
photon/.style={decorate, 
    decoration={snake}},
    math/.style={draw, execute at begin node={$\displaystyle}, execute at end node={$}}
 }

\title[Subhalo properties revealed with high-resolution and large-volume cosmological simulations]{$\Lambda$CDM halo substructure properties revealed with high-resolution and large-volume cosmological simulations}

\author[A. Molin\'e et al.]{\'Angeles Molin\'e$^{1, 2, 3}$
\thanks{E-mail: angeles.moline@upm.es},
Miguel A. S\'anchez-Conde$^{1, 2}$
\thanks{E-mail: miguel.sanchezconde@uam.es},
Alejandra Aguirre-Santaella$^{1, 2}$
\thanks{E-mail: alejandra.aguirre@uam.es},
\newauthor Tomoaki Ishiyama$^{4}$,
Francisco Prada$^{5}$, 
Sof\'ia A. Cora$^{6,7}$,
Darren Croton$^{8,9}$,
Eric Jullo$^{10}$,
\newauthor R. Benton Metcalf$^{11,12}$,
Taira Oogi$^{4,13}$,
Jos\'e Ruedas$^{5}$.
\\
$^{1}$ Instituto de F\'isica Te\'orica UAM-CSIC, Universidad Aut\'onoma de Madrid, C/ Nicol\'as Cabrera, 13-15, 28049 Madrid, Spain\\
$^{2}$ Departamento de F\'isica Te\'orica, M-15, Universidad Aut\'onoma de Madrid, E-28049 Madrid, Spain
\\
$^{3}$ SPace and AStroparticle Group (SPAS), UAH, Madrid, Spain
\\
$^{4}$ Institute of Management and Information Technologies, Chiba University, Chiba, 263-8522, Japan 
\\
$^{5}$ Instituto de Astrof\'isica de Andaluc\'ia (CSIC), Glorieta de la Astronom\'ia, E-18080 Granada, Spain
\\
$^{6}$ Instituto de Astrof\'isica de La Plata (CCT La Plata, CONICET, UNLP), Observatorio Astron\'omico, 
Paseo del Bosque, B1900FWA La Plata, Argentina \\
$^{7}$ Facultad de Ciencias Astron\'omicas y Geof\'isicas, Universidad Nacional de La Plata, Observatorio Astron\'omico, 
Paseo del Bosque, B1900FWA La Plata, Argentina
\\
$^{8}$ Centre for Astrophysics \& Supercomputing, Swinburne University of Technology, Hawthorn, VIC 3122, Australia \\
$^{9}$ ARC Centre of Excellence for All Sky Astrophysics in 3 Dimensions (ASTRO 3D)
\\
$^{10}$ Aix Marseille Univ, CNRS, CNES, LAM, Marseille, France 
\\
$^{11}$ Dipartimento di Fisica \& Astronomia, Universit\`a di Bologna, via Gobetti 93/2, 40129 Bologna, Italy 
\\
$^{12}$ INAF-Osservatorio Astronomico di Bologna, via Ranzani 1, 40127 Bologna, Italy 
\\
$^{13}$ Research Center for Space and Cosmic Evolution, Ehime University, 2-5, Bunkyo-cho, Matsuyama, Ehime 790-8577, Japan 
}

\date{\today}

\pubyear{2020}

\begin{document}
\label{firstpage}
\maketitle

\begin{abstract}
In this work, we investigate the structural properties, distribution and abundance of $\Lambda$CDM dark
matter subhaloes using the Phi-4096 and Uchuu suite of N-body cosmological simulations.
Thanks to the combination of their large volume, high mass resolution and superb statistics,
we are able to quantify -- for the first time consistently over more than seven decades in ratio of  subhalo-to-host-halo mass -- dependencies of subhalo properties on mass, maximum circular
velocity, $V_{\rm max}$, host halo mass and distance to host halo centre. We also dissect the evolution
of these dependencies over cosmic time. We provide accurate fits for the subhalo mass and
velocity functions, both exhibiting decreasing power-law slopes
and with no significant dependence on redshift. We also find subhalo abundance to depend
weakly on host halo mass. Subhalo structural
properties are codified via a concentration parameter, $c_{\rm V}$, that does not depend on any pre-defined density profile and relies only on $V_{\rm max}$. We derive the $c_{\rm V}-V_{\rm max}$ relation and find an important dependence on distance of the subhalo to the host
halo centre. Interestingly, we also find subhaloes of the same mass to be significantly more concentrated
when they reside inside more massive hosts. Finally, we investigate the redshift evolution of
$c_{\rm V}$, and provide accurate fits. Our results
offer an unprecedented detailed characterization of the subhalo population, consistent over a
wide range of subhalo and host halo masses, as well as cosmic times. Thus, we expect our
work to be particularly useful for any future research involving dark matter halo substructure.

\end{abstract}

\begin{keywords}
galaxies: haloes -- methods: numerical -- cosmology: theory -- dark matter
\end{keywords}



\section{Introduction}
\label{sec:intro}

In the current standard model of cosmology, $\Lambda$CDM~\citep{Planck18}, the structure of the Universe is formed via a hierarchical, bottom-up scenario with small primordial density perturbations growing to the point where they collapse into the filaments, walls and eventually dark matter (DM) haloes that form the underlying large-scale-structure filamentary web of the Universe (see, e.g.~\citealt{Frenk12}). Galaxies are embedded in these massive, extended DM haloes teeming with self-bound substructure, the so-called subhaloes. The growth of cosmic structure begins early, where the first haloes to collapse and virialise are smooth triaxial objects. 
These primordial protohaloes continue to grow, and more massive haloes are subsequently formed via merging and accretion of smaller haloes, giving rise to a wide spectrum of halo masses, from the mentioned tiny values up to the most massive haloes in the Universe today, with masses larger than 10$^{15}~\mathrm{M_{\odot}}$.

Numerical simulations 
have proven to be crucial for understanding structure formation in the Universe (see, e.g. \citealt{Kuhlen:2012,Frenk12}). 
Nevertheless, these numerical efforts only cover a limited range in halo masses and redshifts \citep{Ando2019, Zavala:2019}. Resolving small-scale structures is extremely challenging, as the range of lengths, masses, and timescales that need to be simulated is immense. In addition, an expensive computational effort is required to generate at the same time massive haloes with large statistics. This poses a serious challenge for studies particularly focused on the smallest scales. This is the case not only for purely cosmological questions, such as those investigating the dark energy, for which galaxies are considered as cosmological tracers in different volumes of the Universe (see, e.g. \citealt{Amendola:2018}); but also of studies in the field of astroparticle physics, where a large effort is being made to elucidate the nature of the DM particle, that could be made explicit via an understanding of the minimum halo mass, e.g. \citet{Fermi:2015}.

A consequence of the $\Lambda$CDM~ structure formation scenario is the existence of abundant substructure (or subhaloes) within haloes. Studying  the complicated dynamics of these subhaloes within their hosts requires numerical simulations. Unfortunately, state-of-the-art N-body cosmological simulations are not able to resolve the whole subhalo hierarchy. Indeed, being limited by numerical resolution, these simulations typically simulate subhaloes of at least one million solar masses, i.e. orders of magnitude above the minimum halo mass expected in many DM scenarios, and focus on a particular host halo mass scale, like the Milky Way \citep{Diemand08,Springel08}. Furthermore, the finite numerical resolution limiting such simulations implies that a fraction of the subhaloes will be artificially destroyed. As a result, some basic properties of the subhalo population remain uncertain, despite being a fundamental probe of the underlying cosmological model. For example, it is unclear if small subhaloes survive to the intense tidal forces they are subject to from their accretion times to the  present and, if so, under which conditions and orbital configurations \citep{Hayashi_2003,garrisonkimmel17,vandenbosch18,vandenbosch&ogiya18,errani&penarrubia20,grand&white20}. In addition to representing a cosmological test by themselves, understanding both the statistical and structural properties of subhaloes plays a key role for many other diverse studies, such as gravitational lensing \citep{vegetti10}, stellar streams \citep{yoon11,erkal16,bonaca19} and indirect or direct DM detection experiments \citep{masc11,2015JCAP...09..008F,coronado-blazquez19,2019JCAP...12..013I}.

In this work, we improve upon previous studies aimed at characterizing the subhalo population~\citep{RP:2016,Gao:2011,Diemand07}, by making use of data at different cosmic times from the Phi-4096 and Uchuu suite of high-resolution N-body cosmological simulations~\citep{Ishiyama2021}. More precisely, the superb numerical resolution and halo statistics of these simulations allow for a careful and dedicated study of the dependency of subhalo abundance on halo host mass as a function of subhalo mass, maximal circular velocity and distance to the host halo centre. In addition, the structural properties of subhaloes, codified by subhalo concentration, can be studied in detail as well. \citet{Moline17} investigated  subhalo concentrations over several orders of magnitude of subhalo mass for Milky Way-sized systems. In this work, we extend such previous analyses to subhaloes inhabiting host haloes with very different masses, $10^7~h^{-1}\, \mathrm{M_{\odot}} \lesssim M_{\rm h} \lesssim 5 \times 10^{15}\, h^{-1}\, \mathrm{M_{\odot}}$. This will allow us not only to explore subhalo concentrations up to higher subhalo maximal circular velocities, but to also include the dependencies on host halo mass. Furthermore, no evolution of subhalo concentrations with cosmic time was studied in \citet{Moline17} and it is indeed very scarce in the literature as of today~\citep{Emberson:2015,Ishiyama:2020}. Such study will be presented here up to $z=4$ with superb statistics and over a large range of subhalo and host halo masses. 

The work is organized as follows. In Section~\ref{sec:simu} we provide an overview of the simulations we use and summarize their most relevant parameters. In Section~\ref{sec:shprop} we present our comprehensive study of subhalo abundances, radial distribution and concentrations. Section~\ref{sec:redshift} is devoted to the characterization of the mentioned subhalo properties with cosmic time. Finally, we summarize our findings and discuss them in Section~\ref{sec:conc}.


\section{Simulations}
\label{sec:simu}


We use three large cosmological N-body simulations, and their basic
properties are listed in Tab.~\ref{tab:simulation}.  
Two of them, Uchuu and ShinUchuu, 
comprise the Uchuu simulation suite~\citep{Ishiyama2021}. 
The Uchuu simulation consists of
$12800^3$ dark matter particles covering a comoving box of side length
2.0 $h^{-1} \rm Gpc$, with resulting mass resolution of $3.27 \times 10^8\, 
h^{-1}\, \mathrm{M_{\odot}}$.  ShinUchuu is a higher resolution
simulation with $6400^3$ particles covering a 140 $h^{-1} \rm Mpc$
side length box, with resulting mass resolution of $8.97 \times 10^5 \, h^{-1}\,
\mathrm{M_{\odot}}$.  The remaining simulation refers to a small box but with an extremely high resolution, Phi-4096~\citep{Ishiyama2021}. This simulation uses $4096^3$ particles
covering a 16 $h^{-1} \rm Mpc$ side length box, resulting in a mass resolution of $5.13
\times 10^3 \, h^{-1}\, \mathrm{M_{\odot}}$.
Tab.~\ref{tab:All_minmax-subh} shows the mass ranges that we have used for this work. The subhalo (halo) mass $m_{\rm vir}$ ($M_{\rm h}$), is defined as the mass contained within the subhalo (halo) virial radius at $z=0$.  A resolution cut has been applied to every simulation in order to get rid of those hosts with less than 700 particles. This value has been chosen after performing the necessary checks to find the smallest values that we can safely use in each case, i.e. once we have verified the (lack of) resolution effects for different values of the minimum number of particles. 

The cosmological parameters of the Uchuu simulations are
$\Omega_0=0.3089$, $\Omega_{\rm b}=0.0486$, $\lambda_0=0.6911$,
$h=0.6774$, $n_{\rm s}=0.9667$, and $\sigma_8=0.8159$.  Those of the
Phi-4096 simulation are $\Omega_0=0.31$, $\Omega_{\rm b}=0.048$,
$\lambda_0=0.69$, $h=0.68$, $n_{\rm s}=0.96$, and $\sigma_8=0.83$.
Both parameter sets are consistent with the latest measurement by the
Planck satellite~\citep{Planck2020}, 
although they are slightly different from each other.
The largest difference is $\sigma_8$: 
1.6 per cent larger in the Uchuu simulations. 
Such a small difference has a negligible impact on the average concentration 
of a halo~\citep[e.g. ][]{Dutton2014} and, thus, we will treat the three simulations as if the cosmology was exactly the same. 

To find gravitationally bound haloes and subhaloes within the particle
data, the
\textsc{rockstar}\footnote{\url{https://bitbucket.org/gfcstanford/rockstar/}}
phase space halo/subhalo finder~\citep{Behroozi2013} was applied. Their mass and maximum circular velocity are instantaneous at $z=0$ and are calculated using only bound particles. The
halo and subhalo catalogs and their merger trees constructed by
\textsc{consistent trees}~\footnote{\url{https://bitbucket.org/pbehroozi/consistent-trees/}}
code~\citep{Behroozi2013b} are available on the Skies \& Universes
site.~\footnote{\url{http://skiesanduniverses.org/}.}  

\begin{table}
\centering
\begin{tabular}{lcccc}
\hline
\shortstack{Name \\ {}}  & \shortstack{$N$ \\ {}} & \shortstack{$L$ \\($h^{-1}  \rm Mpc$)} &  
\shortstack{$m$ \\ ($h^{-1}\, \mathrm{M_{\odot}}$)} & \shortstack{$\varepsilon$ \\ ($h^{-1} \rm kpc$)}\\
 \hline 
Uchuu & $12800^3$ & 2000 & $3.27 \times 10^{8}$ & $4.27$  \\ 
ShinUchuu & $6400^3$ & 140 & $8.97 \times 10^{5}$ & $0.40$  \\ 
Phi-4096 & $4096^3$ & 16 & $5.13 \times 10^{3}$ & $0.06$  \\ \hline 
\end{tabular}
\caption{
Main properties of the simulations used in this work. 
Here, $N$, $L$, $m$, and $\varepsilon$ are the total number of particles, box
side length, particle mass resolution, and softening length, respectively.
}
\label{tab:simulation}
\end{table}

\begin{table*} 
	\begin{center}
		\begin{tabular}{| l c c c c | }
			\hline
	    	& $V_{\rm max}^{\rm min},  V_{\rm max}^{\rm max}$ &   log$_{10}$ $m_{\rm vir}^{\rm min},$ log$_{10}$ $m_{\rm vir}^{\rm max}$ & $V_{\rm max,h}^{\rm min}, V_{\rm max,h}^{\rm max}$  &   log$_{10}$ $M_{\rm h}^{\rm min},$ log$_{10}$ $M_{\rm h}^{\rm max}$\\ \hline 
			Phi-4096 & [1.0, 234.9] & [4.0, 12.0] & [4.5, 371.4] & [6.6, 12.6]\\ 
			ShinUchuu   & [4.0, 725.7]  & [6.3, 13.7] & [12.5, 842.2] & [8.8, 13.8] \\ 
			Uchuu & [35.0, 1874.4] & [8.8, 15.2] & [89.0, 2582.1] & [11.4, 15.7]\\ \hline 
		\end{tabular}
	\end{center}
	\caption{Minimum and maximum values of maximum circular velocities $V_{\rm max}$ in km s$^{-1}$, and masses in log$_{10}$ [$m_{\rm vir}$/($h^{-1}$ M$_{\odot}$)] for both subhaloes (first two columns) and hosts (third and fourth columns, $V_{\rm max,h}$ [km s$^{-1}$] and log$_{10}$ [$M_{\rm h}$/($h^{-1}$ M$_{\odot}$)]). We have set a minimum of 700 particles for hosts in order to have well resolved data; see text for details.}
	\label{tab:All_minmax-subh}
\end{table*}

\citet{vandenbosch&ogiya18} claimed that cosmological simulations
suffer from significant overmerging due to inadequate force softening,
which causes numerical disruption.  Such numerical disruption can
decrease subhalo abundances in cosmological simulations by up to a
factor of 2~\citep{Green2019}.  Follow-up work by the same author~\citep{Green2021} 
has shown that the effect of numerical disruption is
rather weak compared to the previous claim.  The impact of numerical
disruption was only 10-20 per cent for subhalo abundances,
number density profiles, and substructure mass fractions.  In
cosmological simulations, the mass resolution rather than numerical
disruption is the primary limitation when studying subhaloes.

These and other works~\citep{errani&penarrubia20, 2021arXiv211101148A} 
suggest that numerical disruption
has only a minor effect on subhalo statistics, and a traditional
resolution cut (e.g. minimum number of particles) is more important
to make robust predictions.  Therefore, we perform a
traditional resolution test in 
this work and do not take the effect of
numerical disruption into account.

\section{Subhalo properties}
\label{sec:shprop}

In this section, we will characterize some of the main substructure properties as found in the simulations, namely the subhalo mass function (SHMF) and velocity function (SHVF), the radial distribution within the hosts (SRD) and the subhalo concentrations.

\subsection{Abundances}
\label{sub-sec:abundances}



The cumulative SHMF is usually approximated by a power law: 

\begin{equation}
N (> m_{\rm vir}) = c\, \left( \frac{m_{\rm vir}}{M_{\rm h}} \right)^{-\alpha} \,,
\end{equation}
where $M_{\rm h}$ is the mass of the host halo, $m_{\rm vir}$ is the mass of the subhalo, $c$ is a constant and $\alpha$ is the slope, which according to simulations is found to be in the range $0.9-1$~\citep{Springel08,Diemand07}, a value that agrees well with theoretical expectations~\citep{Gioc1, BlancLaval}. Some studies also suggest that this slope may depend on the host halo mass~\citep{Hellwing:2015upa}.
This is indeed in agreement with results from semi-analytical models (e.g.~\citealt{2018PhRvD..97l3002H}). 

Yet, in practice the SHMF is not perfectly fitted by a power law in the entire range covered by one specific simulation, since it declines more rapidly at the largest masses (i.e. there is no substructure with mass larger than a significant, $\mathcal{O}$(0.1) fraction of the mass of the host), and it decreases at small masses as well because of numerical resolution effects.\footnote{See Section~\ref{sec:simu} for further discussion.
} 

In this regard, a better description of the SHMF is~\citep{RP:2016}:   
\begin{equation} \label{eq:rp16m}
    f = (x/\mu)^{-\alpha}  e^{-(x/\mu_{\rm cut})^{\beta}},
\end{equation}
where $x = m_{\rm vir}/M_{\rm h}$, $\alpha$ represents the slope of the power law and there are three other free parameters. This expression captures the correct behaviour at the high mass end.

As subhaloes lose mass due to tidal stripping when orbiting around the host, the mass parameter for subhaloes, as usually defined with respect to a given overdensity value, may not be valid and can be considered ill-defined. Instead, the tidal mass is 
sometimes used, i.e. the enclosed mass within the tidal radius, or the radius of the subhalo after its interaction with the host tidal forces~\citep{2008gady.book.....B, vandenbosch18}.
In the case of the \textsc{ROCKSTAR} halo/subhalo finder used in this paper, 
the subhalo mass is defined with respect to a given overdensity value 
applied to only particles gravitationally bound to the subhalo, ameliorating ambiguity in the mass definition.

In this work, to avoid complications with the definition of subhalo masses, we will mainly use the $V_{\rm max}$ parameter instead, representing the maximum circular velocity of particles within the subhalo. This quantity is more reliable since it is much less affected by tidal forces~\citep{Kravtsov04, Diemand07}, and allows us to describe the structural properties of a subhalo independently of any density profile and of the particular definition used for the virial/tidal radius \citep{Moline17}. The cumulative SHVF follows a power law as well:
\begin{equation} 
\label{eq:plvmax}
N (> V_{\rm max}) = d\, \left( \frac{V_{\rm max}}{V_{\rm max,h}} \right)^{-\alpha} \,,
\end{equation}
where $V_{\rm max,h}$ is the maximum circular velocity of the host halo, $d$ is a free parameter and $\alpha$ is the slope, which in this case is close to 3~\citep{Springel08, Diemand07, BolshoiP16, RP:2016}. 
The SHVF also experiences the same deviations described for the SHMF at the high and low-mass end, and can also be described using an expression similar to Eq.~\ref{eq:rp16m}, where $x = V_{\rm max}/V_{\rm max,h}$.

%
\begin{figure}
\includegraphics[width=\columnwidth]{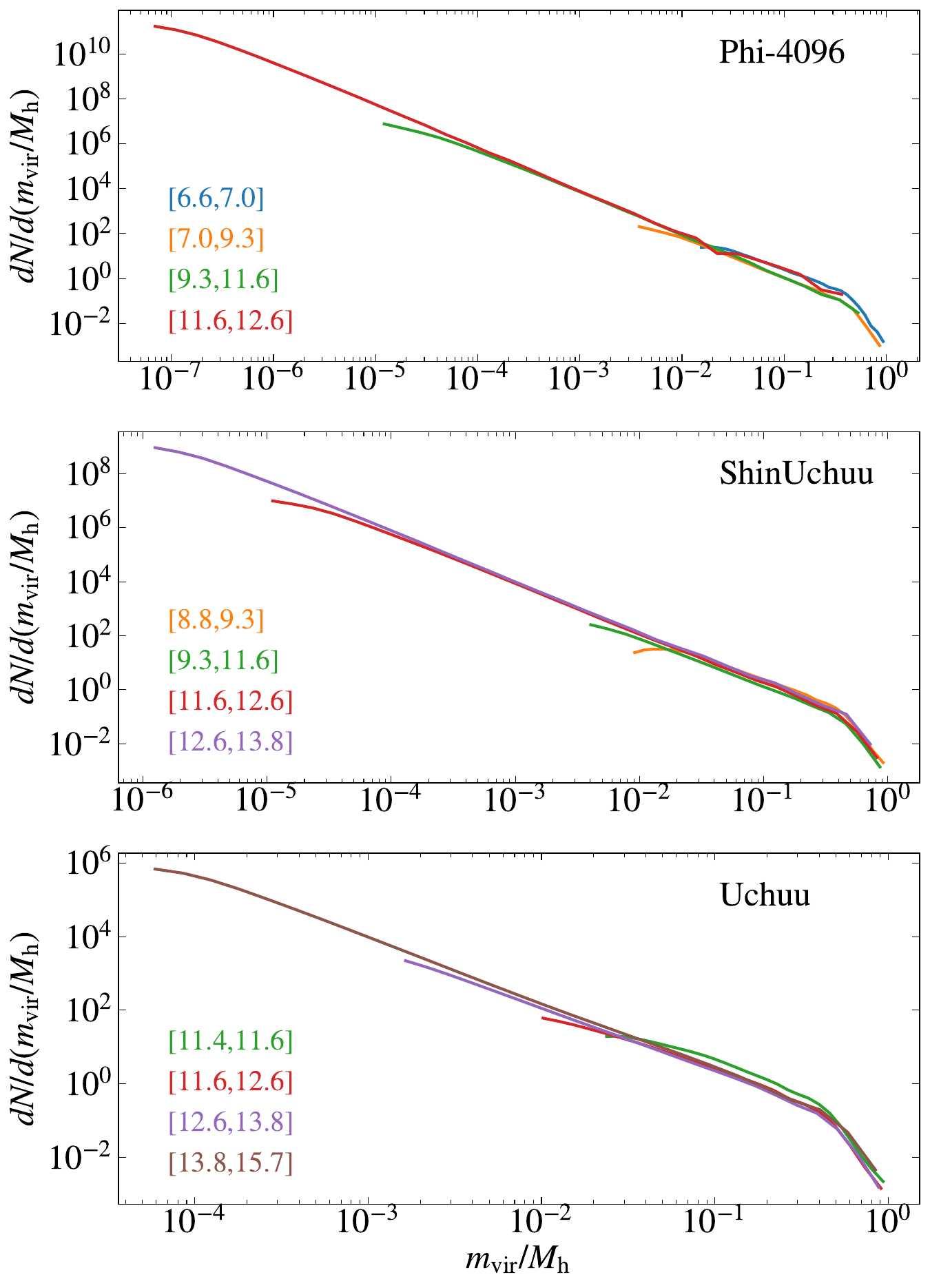}
\caption{
From top to bottom, SHMF of Phi-4096, ShinUchuu and Uchuu at $z = 0$. In each case, the dataset has been divided in four according to the host halo mass in log$_{10}$ [$M_{\rm h}$/($h^{-1}$ M$_{\odot}$)], depicted with different colours in the legend. }
\label{fig:shmfzeroale}
\end{figure}

\begin{figure}
\includegraphics[width=\columnwidth]{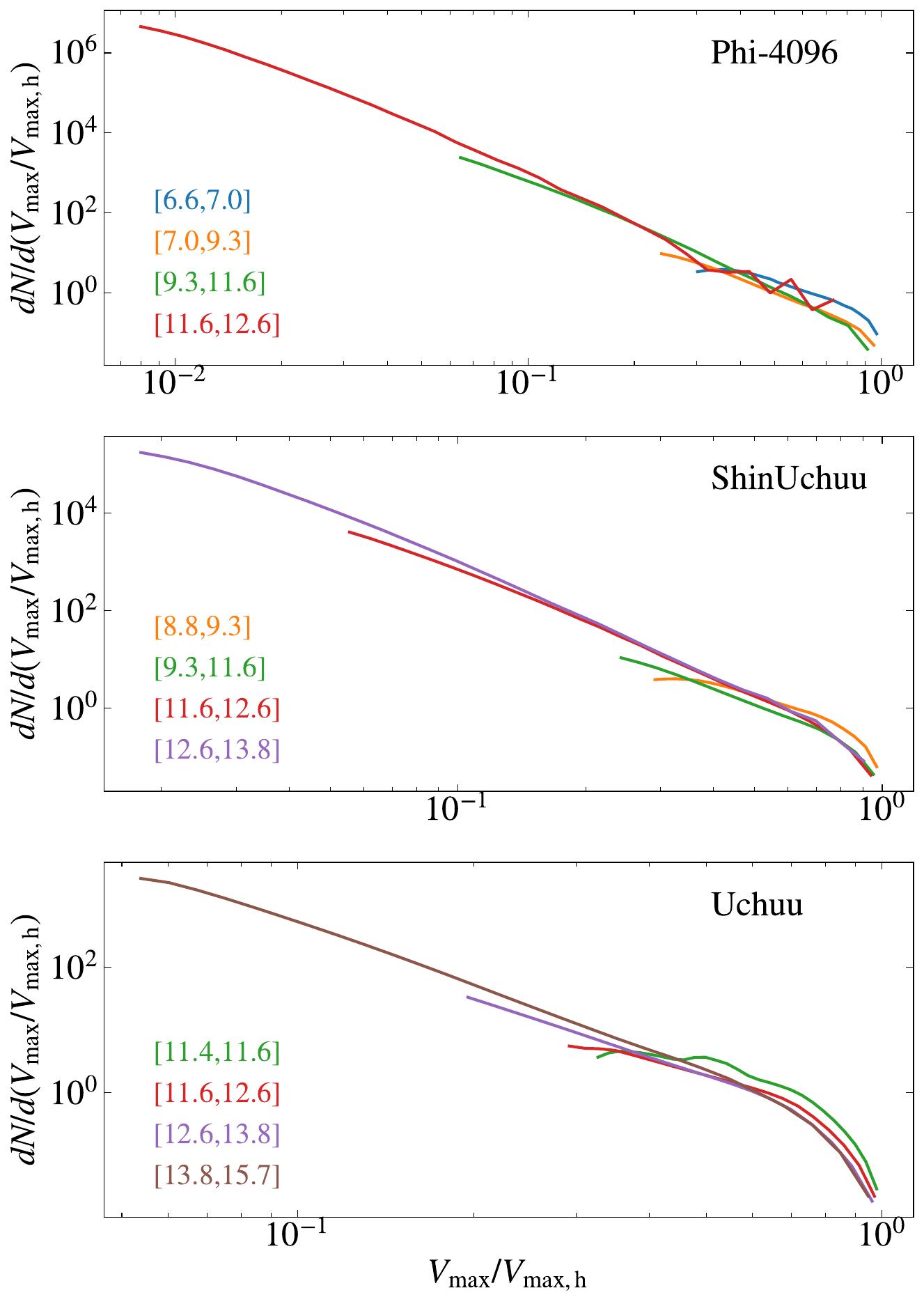}
\caption{
From top to bottom, SHVF of Phi-4096, ShinUchuu and Uchuu, at $z = 0$. In each case the dataset has been divided in four according to the host halo mass in log$_{10}$ [$M_{\rm h}$/($h^{-1}$ M$_{\odot})$], depicted with different colours in the legend.}
\label{fig:shvfzeroale}
\end{figure}
In our analyses, we split the whole dataset into different $m_{\rm vir}$ and $V_{\rm max}$ bins with respect to the host halo mass, and use the ratio between the subhalo mass and its host mass, which will allow for a more optimal comparison between cases. Fig.~\ref{fig:shmfzeroale} shows the SHMFs of Phi-4096, ShinUchuu and Uchuu for several host halo mass bins, averaging over all the hosts in each.
We explain the criteria for our choice to define the mass ranges used in this work in Section~\ref{sub-sec:concentrations}.

 Note that, for larger host masses, the SHMF is resolved down to smaller fractions of the host halo mass. In particular, in the best cases we can resolve down to ratios $\sim10^{-6}$. The same occurs for the SHVFs, shown in Fig.~\ref{fig:shvfzeroale} considering the ratio between $V_{\rm max}$ and $V_{\rm max,h}$. In this case, we obtain the presumed power-law behaviour for ratios approaching $10^{-2}$ at best. For both the SHMF and SHVF, the expected drop at ratios close to 1 is clearly visible in the corresponding figures, as well as the flattening at the low-mass and low-velocity ends due to resolution effects. The overlap between different host halo masses in each simulation is quite accurate (the differences lie below a factor~$\sim$2) for both the SHMF and the SHVF, though some differences can be perceived, especially at the largest subhalo-to-halo ratios. 

\begin{figure}
\centering
\includegraphics[width=\columnwidth]{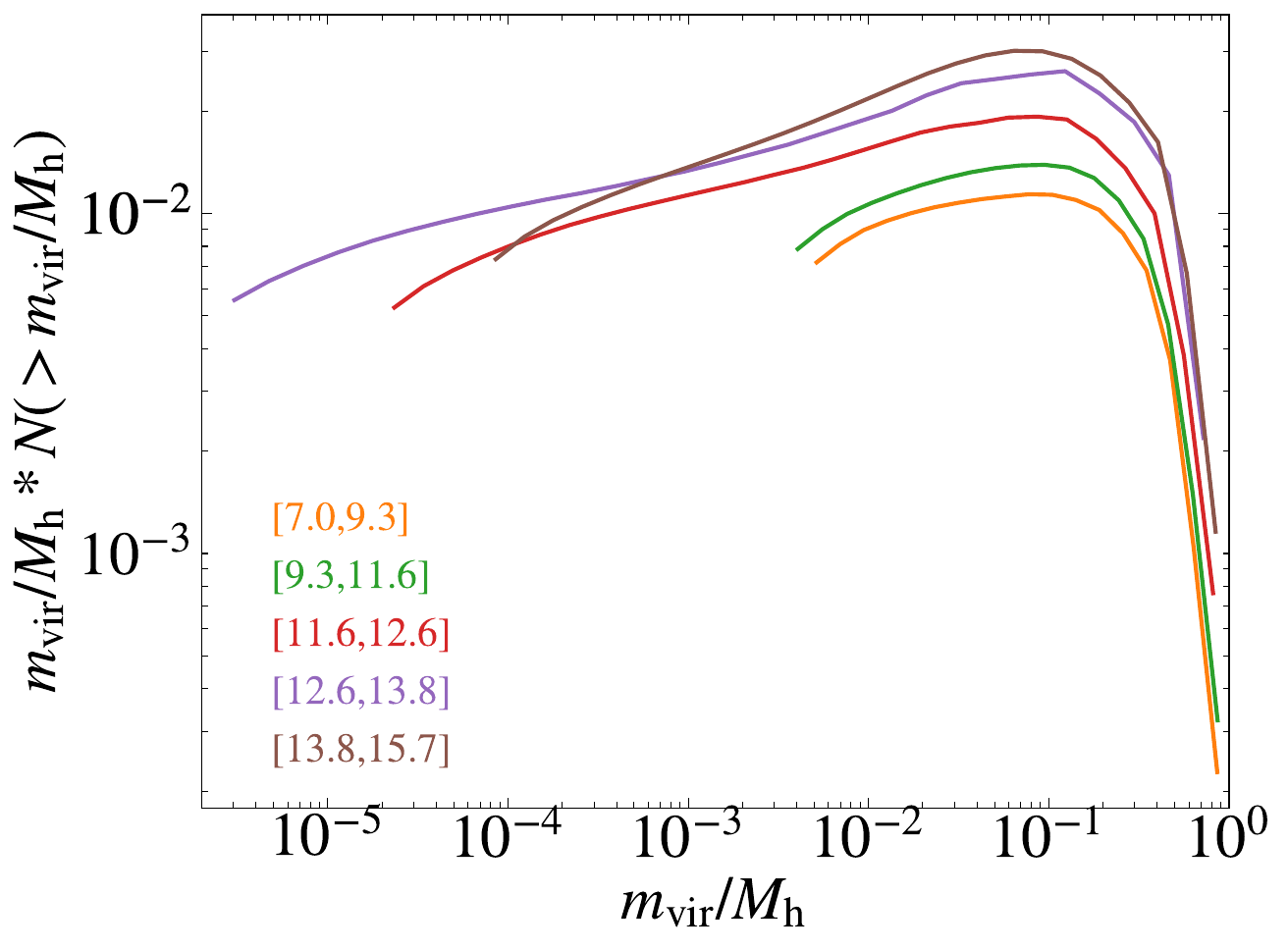}
\caption{Cumulative SHMF for different host halo masses, multiplied by the $x$ axis, i.e. the ratio between the subhalo mass and the host mass in log$_{10}$ [$M_{\rm h}$/($h^{-1}$ M$_{\odot}$)], as indicated with different colours in the legend. The Phi-4096 simulation has been used for the smallest bin, in orange; the next three bins, in green, red and purple, are obtained from ShinUchuu; the brown bin corresponds to Uchuu. }
\label{fig:shmfmwalegao}
\end{figure}

The dependence of subhalo abundance on host halo mass is shown in Fig.~\ref{fig:shmfmwalegao}, where we multiply the cumulative SHMF by the $x$ axis, i.e. the ratio between subhalo mass and host halo mass, in order to visually increase such dependency. 
In this figure, we show results from the largest host halo mass bin in Uchuu, the three well-resolved, intermediate mass ones in ShinUchuu and the smallest well-resolved bin in Phi-4096, i.e. for each host halo mass we choose the most appropriate simulation. 
Interestingly, we find more subhaloes as we increase the parent mass. In particular, we find a factor of 2.6 difference between the SHMF of the largest considered parent mass and the smallest at $m_{\rm vir}/M_\mathrm{h} = 0.1$. This behaviour, already described, e.g. in~\citet{Gao:2011, Ishiyama2013,BolshoiP16}, is now confirmed and quantified to a greater detail in our study.  

\begin{figure}
\centering
\includegraphics[width=\columnwidth]{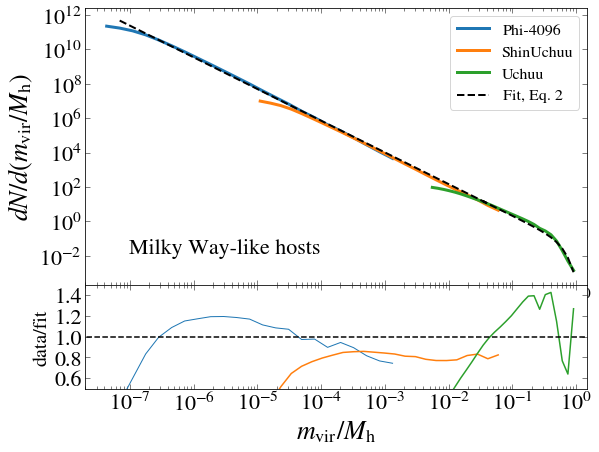}
\includegraphics[width=\columnwidth]{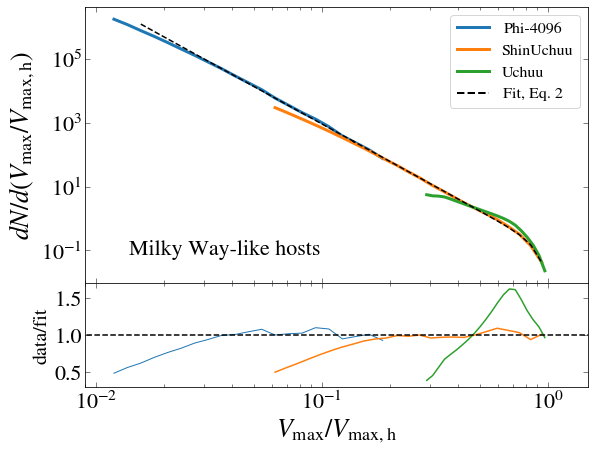}
\caption{SHMF and SHVF (upper and lower panel, respectively)  built from Milky Way-like haloes, with masses between 10$^{11.6-12.6}~h^{-1}~\mathrm{M}_\odot$, in all three simulations at once, each represented with a different colour according to the legend. Our best-fitting using the parametric function by~\citet{RP:2016}, and reproduced in the Eq.~\ref{eq:rp16m}, is also shown in both panels as a dashed line. At the bottom of each panel, the difference between data and model is also shown.}
\label{fig:shmfmwale}
\end{figure}

In Fig.~\ref{fig:shmfmwale}, we focus on Milky Way-like hosts using all simulations at once 
to find the SHMF and SHVF best fits when described with the parametric function of \citet{RP:2016}, and reproduced in the Eq.~\ref{eq:rp16m}. 
To perform the corresponding fits, we consider the regions of the SHMF and SHVF where we find either a power-law (towards low subhalo masses) or an abrupt decay (at the high-mass end). The technical tool to perform such fits was the Python built-in function \texttt{curve\_fit} from \texttt{scipy}. 
In particular, the adopted subhalo mass ranges in units of the host halo mass are $[7.24 \cdot 10^{-7}, 1.32 \cdot 10^{-3}]$, $[0.0014, 0.041]$ and $[0.035, 0.91]$ 
for Phi-4096, ShinUchuu and Uchuu, respectively; and the corresponding $V_\mathrm{max}$ intervals are $[0.031, 0.14]$ and $[0.14, 0.94]$ 
in units of the host halo $V_\mathrm{max}$ for Phi-4096 and ShinUchuu (the Uchuu range was already covered by ShinUchuu). Our resulting fits are also shown. The found best-fitting parameters are provided in Tab.~\ref{tab:rp16fit}. The agreement among the simulations is noticeable, and the best-fitting function matches reasonably well --for the SHMF, the differences are, at most, $\sim$20 per cent for Phi-4096 and ShinUchuu, and $\sim$40 per cent for Uchuu; for the SHVF, they are less than $\sim$10 per cent for Phi-4096 and ShinUchuu and up to $\sim$50 per cent for Uchuu-- over nearly seven and two orders of magnitude in subhalo-to-host-halo mass and velocity ratios, respectively (see residuals in Fig.~\ref{fig:shmfmwale}). The Uchuu cumulative SHMF as a function of the subhalo mass defined at the time of first accretion and for different host haloes mass bins is presented in~\cite{Ishiyama2021}.  As a reference, they also show the Rodr\'iguez-Puebla parametrization~\citep{RP:2016} using the Bolshoi Planck / MultiDark Planck simulations~\citep{BolshoiP16}. The slope at the low-mass end has no significant dependency on host mass. This is in agreement with our results shown in Fig.~\ref{fig:shmfzeroale}. 
Our slope $\alpha$ values also agree well with those reported in previous works~\citep{Springel08,Diemand07, BolshoiP16, Gioc1, BlancLaval} for both the SHMF and SHVF.

\begin{table} 
	\begin{center}
		\begin{tabular}{| l c c c c | }
			\hline
	    	& $\alpha$  &  $\beta$ &  $\mu$ &  $\mu_{\rm cut}$ \\ \hline 
	    	SHMF & 1.84 &  2.91 &  0.15 &  0.59 \\ 
			SHVF & 3.91 &  9.72 &  0.57 &  0.92  \\ \hline 
		\end{tabular}
	\end{center}
	\caption{Best-fitting parameters to the SHMF and SHVF using the parametric form of Eq.~\ref{eq:rp16m}~\citep{RP:2016}, for the case of MW-size host haloes in all our three sets of simulations. The corresponding fits are illustrated in Fig.~\ref{fig:shmfmwale}.}
	\label{tab:rp16fit}
\end{table}

\subsection{Radial distribution}

We have also studied the distribution of subhaloes within their hosts. In this case, we consider all subhaloes in each simulation at $z=0$, and use 6 logarithmic radial bins within the hosts in terms of $x_\mathrm{sub} = r_\mathrm{sub}/R_\mathrm{vir,h}$, where $r_\mathrm{sub}$ is the location of the subhalo in terms of distance to the host halo centre, and $R_\mathrm{vir,h}$ is the virial radius of the host. 

\begin{figure}
\centering
\includegraphics[width=\columnwidth]{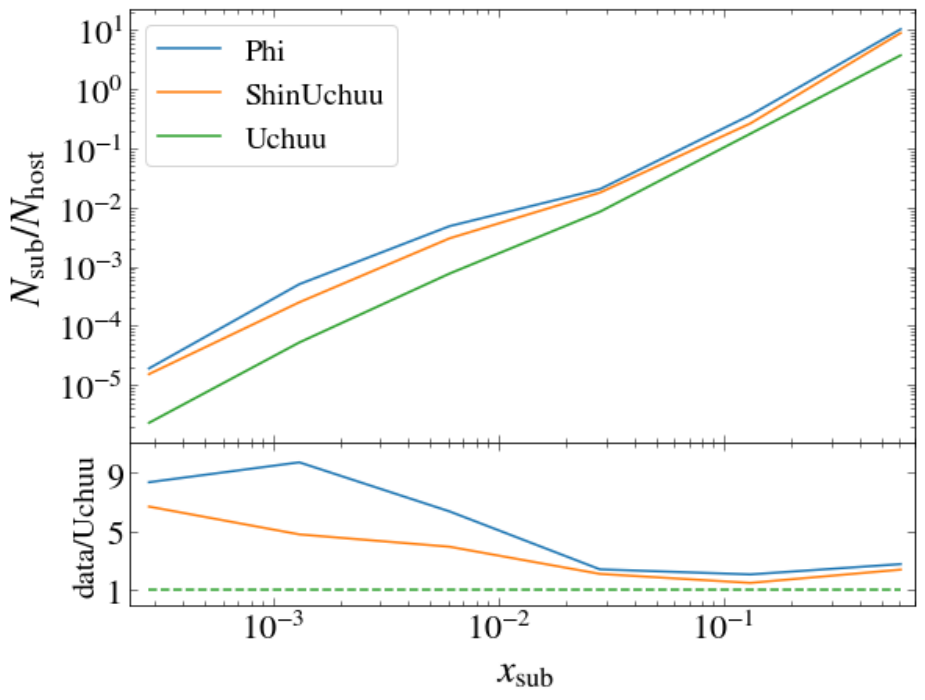}
\caption{SRD of each simulation at the present time, including subhaloes into all host haloes. The $x_{\rm sub}$ parameter is the radial position of the subhalo in units of the virial radius of its host. To compute each SRD, we average over the total number of hosts in that simulation, $N_\mathrm{host}$. Six logarithmic equally spaced radial bins have been used in each case. 
}
\label{fig:srdz0}
\end{figure}

The subhalo radial distributions (SRDs) at the present time are shown in Fig.~\ref{fig:srdz0} for each simulation. 
In this figure, we show the number of subhaloes in each radial bin, divided by the total number of hosts. 
Our SRD results confirm that most subhaloes are located in the outskirts of the host (although the subhalo number density is higher as we approach the centre). Remarkably, we have subhaloes lying inside one thousandth of the virial radius of the host. Also, as expected, a larger number of subhaloes is obtained in the innermost parts of the host for the simulations with better numerical resolution, that is, a smaller minimum subhalo mass. In particular, we find roughly a factor 10 more subhaloes in Phi-4096 than in Uchuu at $x_\mathrm{sub} = 10^{-3}$, and still a factor 5 more subhaloes in ShinUchuu compared to Uchuu. 

\subsection{Concentrations}
\label{sub-sec:concentrations}
There is no consensus today on the most accurate way to describe the DM density profile of subhaloes. Although it is possible to study the distribution of DM particles inside subhaloes using simulations, the innermost region cannot yet be satisfactorily modeled due to numerical resolution (see, e.g. the discussion in~\citealt{Diemand_2011} and~\citealt{Green2019}). In addition, it is well known that tidal stripping removes mass from the outer parts of subhaloes, causing the distribution of DM to fall abruptly there and then the virial radius of subhaloes is not well
defined~\citep{Ghigna98,Taylor:2000zs,Kravtsov04,Kazantzidis04,Diemand:2006ik,Diemand:2006ik,Springel08}. As a consequence, the subhalo concentration cannot follow the formal definition used for halo concentration, $c_{\Delta} \equiv R_{\rm{vir,h}}/r_{-2}$, i.e. the ratio of the halo virial radius, $R_{\rm{vir,h}}$, and the radius $r_{-2}$ at which the logarithmic slope of the DM density profile $\frac{d \log\rho}{d \log\,r}=-2$. 

An alternative way is to define the subhalo concentration independently of the adopted density profile. This can be done, e.g. by expressing the mean physical density, $\bar{\rho}$, attained within the radius corresponding to the maximum circular velocity, $R_{\rm{max}}$, in units of the critical density of the Universe~\citep{Diemand07, Diemand08, Springel08,Moline17}:
\begin{equation}
\label{eq:cv-def}
c_{\rm V}=\frac{\bar{\rho}(R_{\rm{max}})}{\rho_{c}(z)} = 2\left( \frac{V_{\rm{max}}}{H(z) \, R_{\rm{max}}}\right)^{2} \,,
\end{equation}
 where $\rho_{c}(z)$ and $H(z)$ are, respectively, the critical density
 and the Hubble parameter as a function of redshift, $H(z)=H_{0}\,\sqrt{\Omega_{\rm{m},0}(1+z)^3+\Omega_{\Lambda}}\equiv H_{0}\,h(z)$. We note that there exists an easy way to relate this $c_{\rm V}$ with the more familiar $c_{\Delta}$, so that a comparison with halo concentration before subhalo accretion can also be made, see, e.g. \citet{Diemand07,Moline17}. 
 Other important reason to use the $c_{\rm V}$ definition is that $V_{\rm{max}}$ is achieved at a radius $R_{\rm{max}}$ that does not fall within the inner regions subject to resolution problems (for a typical NFW profile (\citealt{Navarro:1995iw,Navarro:1996gj}), for instance, $V_{\rm{max}}$ occurs at $R_{\rm{max}}=2.163 \, r_{-2}$, and the relation does not vary drastically for other profiles).

Tab.~\ref{tab:All_minmax-subh} provides the $V_{\rm max}$ and mass ranges covered by our set of simulations for both host haloes and subhaloes at $z=0$. In order to determine the subhalo concentrations using the definition in Eq.~\ref{eq:cv-def}, we apply  additional, specific cuts on the subhalo maximum circular velocity in order to avoid numerical resolution issues. These cuts are based on that found in the $R_{\rm max} - V_{\rm max}$ parameter space: 
the expected behaviour of the $R_{\rm max} - V_{\rm max}$ relation is almost linear as was studied in other works (see \citealt{Xu:2015,Zavala:2019}).  We avoid the $V_{\rm max}$ values at which this behaviour is no longer fulfilled 
(see Appendix \ref{sec:appendix} for further details).
\begin{table} 
	\begin{center}
		\begin{tabular}{| l c c c c | }
			\hline
	    	& $V_{\rm max}^{\rm min}$  &   log$_{10}$ $m_{\rm vir}^{\rm min}$ &  $V_{\rm max,h}^{\rm min}$ &  log$_{10}$ $M_{\rm h}^{\rm min}$ \\ \hline 
	    	Phi-4096 & 7.0  & 6.4 & 7.0 & 7.0 \\ 
			ShinUchuu   & 38.0   & 8.8 & 40.0 & 9.3\\ 
			Uchuu & 180.0 & 11.0 & 270.0 & 11.6 \\ \hline 
		\end{tabular}
	\end{center}
	\caption{Minimum values of masses in log$_{10}$ [$(m_{\rm vir}/(h^{-1}$ M$_{\odot})$] for subhaloes and haloes, log$_{10}$ [$(M_{\rm h}/(h^{-1}$ M$_{\odot})$]), and their corresponding minimum values of circular velocities in km s$^{-1}$, considered for the study of concentrations in Phi-4096, ShinUchuu and Uchuu at redshift $z=0$. Note that these values differ from those shown in Tab.~\ref{tab:All_minmax-subh} to avoid the impact of resolution effects on $c_{\rm V}$ values; see text for details.}
	\label{tab:Cv-minmax}
\end{table}
After applying these pre-selection cuts on the data, the minimum values of $V_{\rm max}$ used in the determination of $c_{\rm V}$ for each simulation at redshift $z=0$ are presented in Tab.~\ref{tab:Cv-minmax}, together with the corresponding minimum values of the subhalo and halo mass.

In order to carefully study the dependencies of the subhalo concentrations with on both $V_{\rm{max}}$ and distance to the host halo centre, we implemented three radial bins within the virial radius of the host halo, following~\cite{Moline17}. 
\begin{figure*}
\centering
\includegraphics[width=0.8\textwidth]{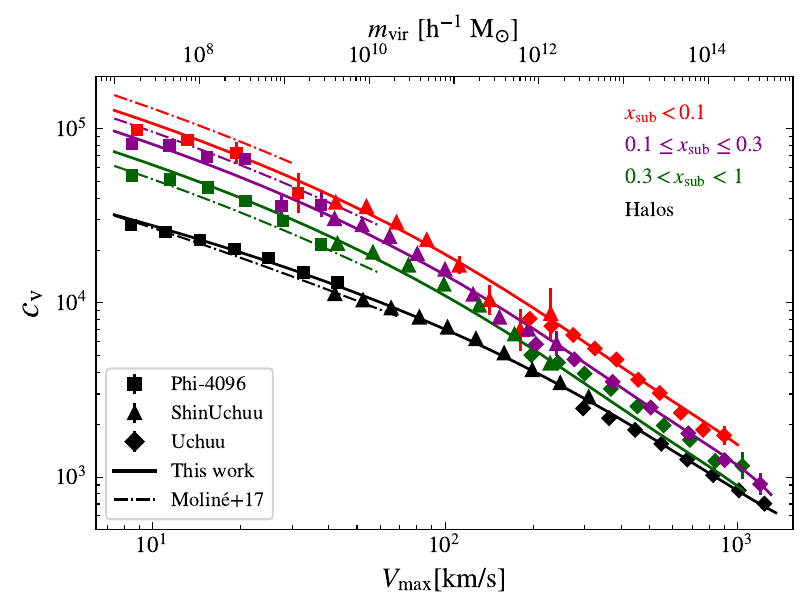}
\caption{Median subhalo and halo concentration parameter $c_{\rm V}$ (Eq.~\ref{eq:cv-def}), as well as standard error of the median as a function of $V_{\rm max}$ (or virial masses, m$_{\rm vir}$, along the x-axis on the top) as found in the Phi-4096 (squares), ShinUchuu (triangles) and Uchuu (diamonds) simulations~\citep{Ishiyama2021}. Both, $V_{\rm max}$ and m$_{\rm vir}$ are directly computed from the simulation data, i.e. no relation was adopted to convert from one to another. Results for subhaloes are shown for three different radial distances to the centre of the host halo. From top to bottom: the innermost bin I (red symbols), intermediate bin II (magenta) and the outermost bin III (green); see figure legend and main text for details. We also include the results for field haloes, represented by black symbols. Solid lines correspond to our fits, both for main haloes (black line) as given by Eq.~\ref{eq:cv-cal}, and for subhaloes (coloured lines) as in Eq.~\ref{eq:cv-fit} for each of the three radial subhalo bins. For comparison, we also show the parametrization in \citet{Moline17} for both field haloes and  subhaloes in similar radial bins (dash-dotted lines).}
\label{fig:cv-Vmax-z0}
\end{figure*}
The innermost radial bin contains subhaloes at a distance $x_{\rm sub} < 0.1$  from the host halo centre (bin I), while the second and third radial bins are defined as $0.1 < x_{\rm{sub}} < 0.3 $ (bin II) and $0.3 < x_{\rm{sub}} < 1$ (bin III), respectively. 

Then, for each radial bin, we grouped subhaloes in bins of $V_{\rm max}$ and  obtained the medians of $c_{\rm V}$. The bin sizes chosen to cover the entire  $V_{\rm max}$ range of each simulation are the same. In Fig.~\ref{fig:cv-Vmax-z0}, we show the median $c_{\rm V}(V_{\rm{max}})$ values and the standard error of the median found for Phi-4096, ShinUchuu and Uchuu. Different colours correspond to the three radial bins, as indicated.  Altogether, they cover the subhalo  maximal circular velocity range between $V_{\rm max} \simeq$ (7$\,-\,1500$) km s$^{-1}$ (or equivalently, $\sim(4 \times 10^6\,-\, 3 \times 10^{14})~h^{-1}$ M$_{\odot}$ in mass). Note that distinct haloes may still overlap and subhaloes are not necessarily fully contained within their hosts. In order to discard such overlapping subhaloes, we only consider those for which their virial radius $r_{\rm vir}$, is fully contained by the virial radius of the host.\footnote{We apply the following condition: $R_{\rm vir,h} > r_{\rm sub}+r_{\rm vir}$. For the Uchuu simulation, we found that $\sim $20 per cent of all subhaloes are overlapping, this value decreasing to $\sim $10 per cent and $\sim $2 per cent in the ShinUchuu and the Phi-4096 simulations, respectively.} 

Remarkably, the figure shows an excellent agreement between the simulations, also in the overlapping $V_{\rm max}$ values. For comparison, we also show the concentration of field haloes obtained using the same definition considered for subhaloes  (Eq.~\ref{eq:cv-def}). As in previous works, we confirm that subhaloes exhibit, on average, higher concentrations than field haloes of the same mass~\citep{Ghigna:1999sn,Bullock:1999he,Ullio:2002pj,Moline17, Ishiyama:2020}. More precisely, we find that c$_{\rm V}$ subhalo values can be up to a factor $\sim$3 larger than those of field haloes of the same $V_{\rm max}$ (for the innermost radial bin and smallest $V_{\rm max}$ of both subhaloes and haloes in Phi-4096 and ShinUchuu), typically being between a factor $\sim$ 1.5 -- 2.5 (its exact number depending on the exact $V_{\max}$ considered and distance to host halo centre). For Uchuu, the ratio between subhalo and halo c$_{\rm V}$ values is typically lower and, indeed, never reaches a factor 2. We conclude that the differences between halo and subhalo concentrations decrease as $V_{\rm max}$ (or, equivalently, the mass) increases.

At this point, it becomes desirable to provide an
approximation that describes the dependence of the median subhalo concentrations on the distance to the host halo centre and the subhalo maximum circular velocity. As in~\cite{Moline17}, we propose a parametrization for the $c_{\rm V}(V_{\rm max}, x_{\rm sub})$ relation, based on the results above:
\begin{eqnarray}
c_{\rm V}(V_{\rm max},x_{\rm sub}) & = & c_{0} \, \left[1+\sum_{i=1}^{3} \,  \left[a_{i} \, \log_{10}\left(\frac{V_{\rm max}}{{\rm km \, s^{-1}}}\right)\right]^{i} \right] \times \nonumber \\ 
& & \left[1 +  b \, \log_{10} \left(x_{\rm sub}\right) \right] ~,
\label{eq:cv-fit}
\end{eqnarray}
where $c_{0} = 1.12 \times 10^5$,  $a_{i}=\left\{\color{blue!70!black}{-0.9512, \, -0.5538, \, -0.3221} \right\}$  and  $b = \color{blue!70!black}{-1.7828}$.

In Fig.~\ref{fig:cv-Vmax-z0} we show the results of this 
fit together with the median concentration values from Phi-4096, ShinUchuu and Uchuu simulations, for all the radial bins considered in our work. The fit works well in the subhalo  $V_{\rm max}$ range $8$~km s$^{-1}$ $\lesssim V_{\rm max} \lesssim 1500$~km s$^{-1}$ 
and the subhalo $x_{\rm sub}$ range  $0.02 \lesssim x_{\rm sub} \lesssim 1.0$, its accuracy being better than 5 per cent at all $V_{\rm max}$ values within this range 
and distances to the host halo centre. For comparison, we also show the \citet{Moline17} parametrization with the dashed lines obtained with data from the VL-II~\citep{Diemand08} and ELVIS~\citep{ELVIS1} N-body simulations. 
We recall that these simulations only describe substructures in MW-like haloes, $\sim$[7 $\times 10^{11} -$ 2 $\times 10^{12}]~h^{-1}$ M$_{\odot}$, and provide subhalo data up to $V_{\rm max} \simeq 30$ km s$^{-1}$ for the innermost radial bin (red symbols), $V_{\rm max} \simeq 60$ km s$^{-1}$ for both the intermediate radial bin (magenta) and the outermost one (green), and $V_{\rm max} \simeq 70$ km s$^{-1}$ for haloes\footnote{In \cite{Moline17}, in order to reduce the uncertainties when extrapolating outside the range probed by the VL-II and ELVIS simulations, the authors used BolshoiP simulation \citep{BolshoiP16} results for more massive {\it haloes}.}; while in this work we are considering {\it all} host haloes (yet with $V_{\rm max}$ values above the corresponding cut) provided by the Uchuu and Phi-4096 simulations. These, we remind, cover a very wide halo mass range, $\sim$[$10^{7} -$ 8.8 $\times 10^{14}]~h^{-1}$ M$_{\odot}$, which explains the observed differences between both the new and the old parametrizations. Our
fit has a similar $V_{\rm max}$ behaviour to that in \cite{Moline17} below $V_{\rm max} \simeq 100$ km s$^{-1}$, however the overall normalisation is different for each radial bin. This is due to the dependence of the subhalo concentration on the host halo mass (as we will see below) when the analysis is performed for different host halo masses and distances to host halo centre. When these dependencies are not taken into account, both fits coincide in the range of subhalo masses ($V_{\rm max}$) covered by the simulations used in \cite{Moline17} (see Section~\ref{subsec:concentration-z}). In this work, we do not include the dependence of the subhalo concentration on the host halo mass in our fit, which will be explored in detail in a future study.
Together, both ShinUchuu and Uchuu simulations cover high $V_{\rm max}$ values ($40$~km s$^{-1}$ $\lesssim V_{\rm max} \lesssim 1500$~km s$^{-1}$), which allows us to provide a fit reaching values more than an order of magnitude above the ones probed in \cite{Moline17} with superb statistics.

Similarly to what has been done for subhaloes in Eq.~\ref{eq:cv-fit}, we obtain a fit for field haloes based also on data from Phi-4096, ShinUchuu and Uchuu (black symbols in Fig.~\ref{fig:cv-Vmax-z0}):
\begin{equation}
c_{\rm V}^{\rm h}(V_{\rm max,h}) = c_{0} \, \left[1+\sum_{i=1}^{3} \,  \left[a_{i} \, \log_{10}\left(\frac{V_{\rm max}}{{\rm km \, s^{-1}}}\right)\right]^{i} \right] ~,
\label{eq:cv-cal}
\end{equation}
where $c_{0} = 7.21 \times 10^4$ and $a_{i}=\left\{-0.81, \, -0.47, \, -0.27 \right\}$. 
\begin{table*}
	\begin{center}
		\begin{tabular}{| l c  c  c  c  c  c | }
			\hline
			 & $c_0$ & $a_1$ & $a_2$ & $a_3$ & $b$ & $d$ \\ \hline
			 $c_{\rm V}(V_{\rm max},x_{\rm sub}) \,$[Eq.~\ref{eq:cv-fit}]   & $1.12 \times 10^5$ & -0.9512 & -0.5538 &  -0.3221 &  -1.7828 & -\\ 
		    $c_{\rm V}(V_{\rm max,h}) \,$ [Eq.~\ref{eq:cv-cal}]   & $7.21 \times 10^4$ & -0.81 & -0.47 & -0.27 &  - & -\\ 
			$c_{\rm V}(V_{\rm max},z) \,$ [Eq.~\ref{eq:cv-z-fit}]   & $1.75 \times 10^5$ & -0.90368 & 0.2749 & -0.028  & -5.52 & 3.2\\ \hline 
		\end{tabular}
	\end{center}
	\caption{Best-fitting values of the parametrizations for the concentration parameter for subhaloes as a function of $V_{\rm max}$ and $x_{\rm sub}$ at $z = 0$ as well as its dependence on $V_{\rm max}$ and redshift. We also provide the parameters of our fit for haloes ($c^{\rm h}_{\rm V}$).}
	\label{tab:fits}
\end{table*}
The fit works well for $7$~km s$^{-1}$ $\lesssim V_{\rm max,h} \lesssim 1500$~km s$^{-1}$, with an error smaller than $\sim$4 per cent within this range for all $V_{\rm max}$ values. In addition, this fit agrees with the one provided in \cite{Moline17} for $V_{\rm max,h} \lesssim 200$~km s$^{-1}$, for which range Phi-4096 and ShinUchuu are used.  However, we find important differences at higher $V_{\rm max,h}$ values already probed by Uchuu.\footnote{Note that, in \cite{Moline17},  $c_{\rm V}^{\rm h}$ values in such an extreme $V_{\rm max}$ range were obtained from $c_{\Delta}$ and not from $V_{\rm max}$ and $R_{\rm max}$; see Eq.~\ref{eq:cv-def}.}
The best-fitting values for the two parametrizations of the concentration described above are indicated in Tab.~\ref{tab:fits}.

\begin{figure*}
\centering
\includegraphics[width=0.49\textwidth]{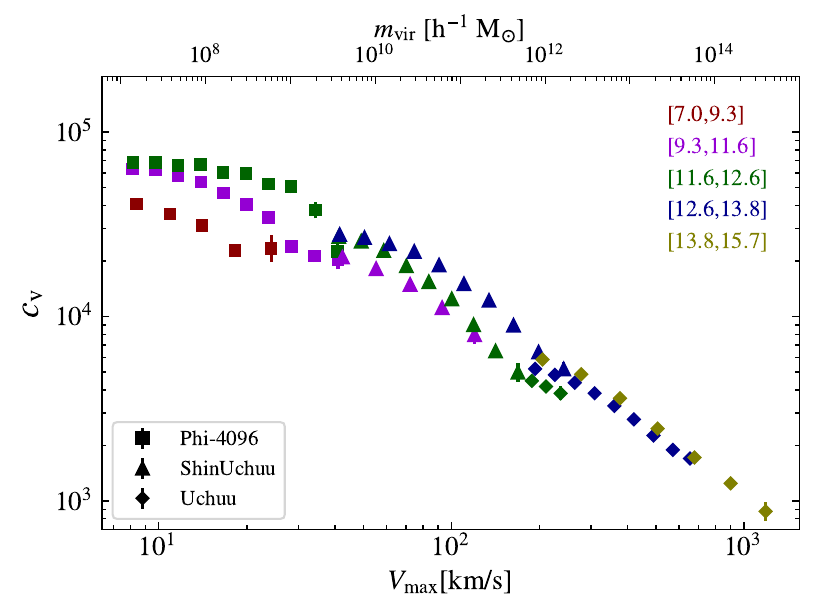}
\includegraphics[width=0.49\textwidth]{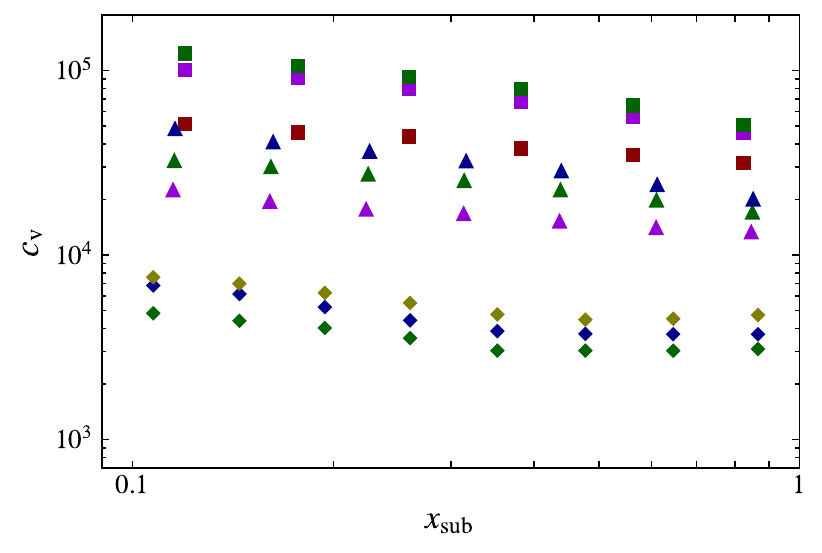}
\caption{Median subhalo concentrations and corresponding standard errors of the median, as found in the Phi-4096, ShinUchuu and Uchuu  simulations. The left panel shows the median $c_{\rm V}$ as a function of $V_{\rm{max}}$ (subhalo virial masses along the x-axis on the top), while the right panel is for $c_{\rm V}$ as a function of $x_{\rm sub}$. We show results for subhaloes residing in different bins of host halo masses (coloured symbols). The values within square brackets in the upper right corners indicate the considered ranges of host halo mass in log$_{10}$ [$M_{\rm h}$/($h^{-1}$ M$_{\odot}$)]. Note that $c_{\rm V}(x_{\rm sub})$ values (right panel) corresponding to the same bin of host halo mass do not need to coincide. They were obtained from different simulations and so, implicitly calculated in a different $V_{\rm{max}}$ range.
}
\label{fig:Cv_xbin_halomass}
\end{figure*}

\begin{figure*}
\centering
\includegraphics[width=0.9\textwidth]{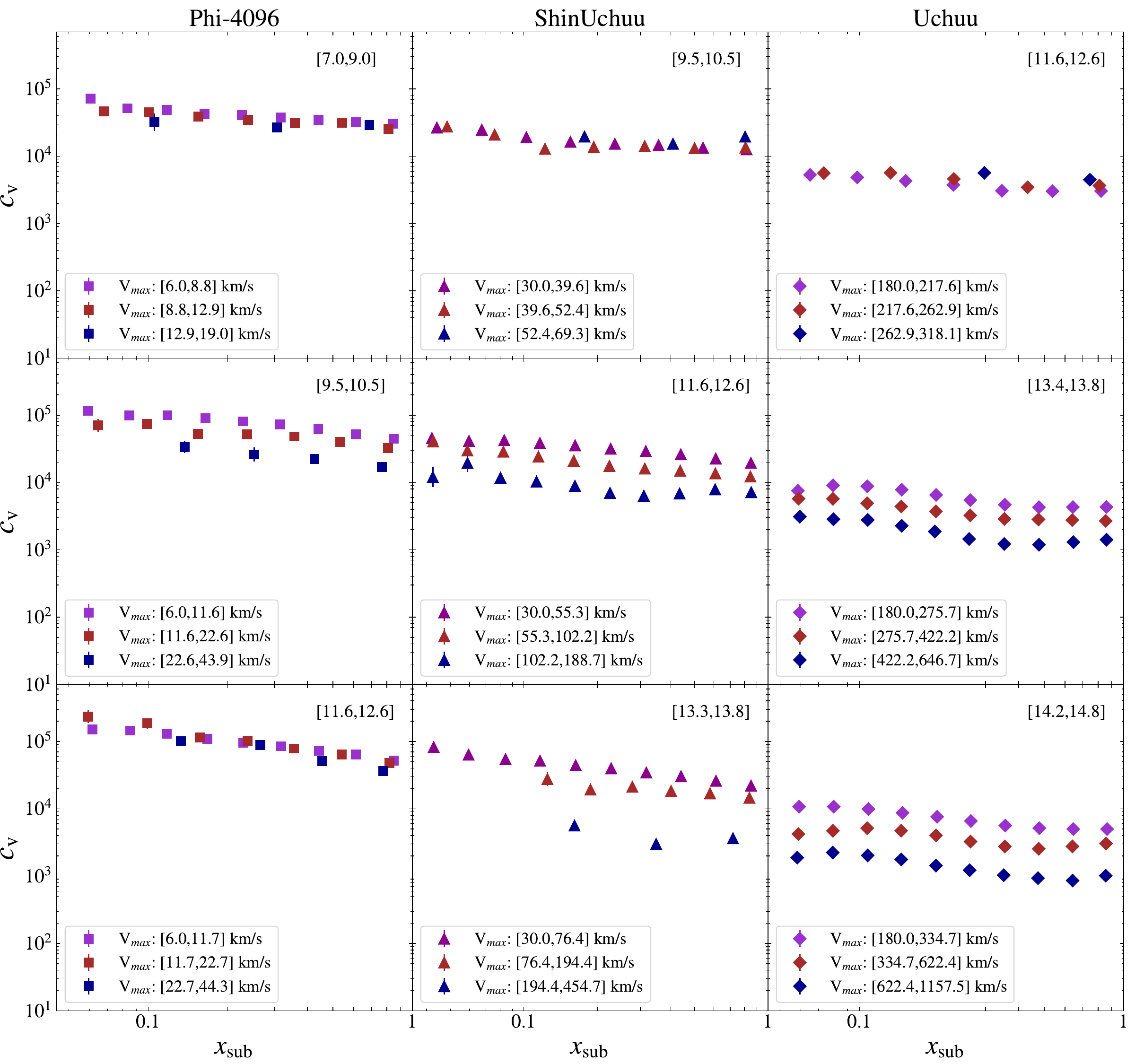}
\caption{Median subhalo parameter $c_{\rm V}$, and the standard error of the median as a function of $x_{\rm sub}$ as found in the Phi-4096 (left column), ShinUchuu (middle column) and the Uchuu (right column) simulations. Each panel shows the results for subhaloes inside a given host halo mass range, as indicated, depicted for three different bins of $V_{\rm max}$.}
\label{fig:Cv_xbin_halomass_vmax}
\end{figure*}

We also paid special attention to the dependence of subhalo concentrations on host halo mass in Phi-4096, ShinUchuu and Uchuu. We present our results in Figs.~\ref{fig:Cv_xbin_halomass} and~\ref{fig:Cv_xbin_halomass_vmax}. In the left panel of Fig.~\ref{fig:Cv_xbin_halomass} we show the medians of $c_{\rm V}$ as a function of  $V_{\rm max}$ for different host halo masses, while the right panel shows $c_{\rm V}$ as a function of $x_{\rm sub}$. Different colours correspond to different bins in host halo mass, as indicated.  
In the figure, host haloes as well as their subhaloes were selected to cover the whole halo mass range provided for each simulation once our pre-selection cuts specified above were applied to the data (see Tab.~\ref{tab:Cv-minmax}). 
The motivation to define the host halo mass ranges used in this work is twofold. On one hand, we wanted to perform our analysis using same $M_{\rm h}$ intervals for each simulation --whenever possible-- in order to have results to compare with between the different simulations. We also wanted to use similar bins throughout the entire paper to both unify and simplify potential comparisons among different sections, thus the host halo mass bins we choose for the study of concentrations correspond to some of those already considered for the study of subhalo abundances (see Section 3.1). On the other hand, we found it convenient to include a mass bin corresponding to MW-like haloes (i.e. log$_{10}$ [$M_{\rm h}$/($h^{-1}$ M$_{\odot}$)] = [11.6,12.6]). 
As we see, the mass ($V_{\rm max}$) ranges used to study the concentrations for each simulation are smaller than those corresponding to subhalo abundances. In this way, we consider the former as a reference to define the host halo mass ranges.
Once we fixed the mass interval corresponding to the MW-like haloes, for Phi-4096 we split the remaining data to study subhalo concentrations into two equal size logarithmic bins (log$_{10}$ [$M_{\rm h}$/($h^{-1}$ M$_{\odot}$)] = [7.0,9.3], [9.3,11.6]). We also adopt these mass bins in both ShinUchuu and Uchuu for the reasons explained above.
The last bin in ShinUchuu corresponds to the most massive host haloes (log$_{10}$ [$M_{\rm h}$/($h^{-1}$ M$_{\odot}$)] = [12.6,13.8]) which was also used for Uchuu. In the latter, still one more bin is possible and necessary to cover the high-mass end, log$_{10}$ [$M_{\rm h}$/($h^{-1}$ M$_{\odot}$)] = [13.8, 15.7]. 
We note that, in the case of subhalo abundances, we included one extra bin for the less massive host haloes in each simulation, i.e. 
log$_{10}$ [$M_{\rm h}$/($h^{-1}$ M$_{\odot}$)] =  [6.6,7.0], [8.8,9.3], [11.4,11.6] for Phi-4096, ShinUchuu and Uchuu, respectively.

Interestingly, we found that, at a given $V_{\rm max}$, subhaloes are systematically more concentrated when they lie inside more massive haloes. Similar results are found when comparing subhalo concentrations as a function of distance from the host halo centre for different host halo mass bins (right panel of Fig.~\ref{fig:Cv_xbin_halomass}).\footnote{In this case, the comparison between $c_{\rm V}$ values should be made individually for each simulation, i.e.  $c_{\rm V}$ values corresponding to the same bin of host halo mass but obtained from different simulations do not need to coincide, as they were implicitly calculated in a different $V_{\rm{max}}$ range (see left panel of Fig.~\ref{fig:Cv_xbin_halomass}). 
}

The origin of this interesting result may be linked to the physical processes that yield the formation and evolution of haloes and their substructure. 
The density perturbations from which haloes form are not independent with on each other. 
During collapse, perturbations are typically affected by the surrounding density field, in such a way that the most massive haloes tend to form in higher density regions and less massive host haloes will form in low density ones~\citep{Doroshkevich:1970,Despali:2013}. Likewise, subhaloes inside these massive haloes would also have a higher ratio between their mean densities and $\rho_{c}(z)$, leading to higher concentrations (see Eq.~\ref{eq:cv-def}) than subhaloes of the same mass --or equivalently, $V_{\rm max}$-- hosted by less massive haloes in less dense environments. In addition, subhaloes inside massive haloes will be affected by stronger tidal disruption at a fixed distance, making them more compact than subhaloes residing in less massive hosts. As a result, the former will have smaller $R_{\rm max}$ than the latter and the enclosed mean subhalo density, codified in $c_{\rm V}$ (Eq.~\ref{eq:cv-def}), increases~\citep{Diemand07,Kuhlen:2008aw,Springel08}. Other explanations may have to do with the most probable orbits subhaloes may take in the most massive hosts compared to the less massive ones (see, e.g.~\citealt{jiang15}). For the moment, though, these are just reasonable conjectures that could explain, at least partially, the results. Further work will be necessary to understand the physical origin of our findings, which will be done elsewhere.

Finally, we extended this study on the dependence of subhalo concentration on host halo mass and distance to the host halo centre by dissecting further the effect considering different $V_{\rm max}$ bins. We show the results in Fig.~\ref{fig:Cv_xbin_halomass_vmax}. 
The five host halo mass bins adopted in this figure were deliberately chosen so that they roughly match the mass scale of dwarf galaxies (log$_{10}$ $M_{\rm h} \simeq [7.0,10]$ $h^{-1}$ M$_{\odot}$), Milky Way-like haloes (log$_{10}$ $M_{\rm h}$ $\simeq [11.5,12.5]~h^{-1}$ M$_{\odot}$), galaxy groups (log$_{10}$ $M_{\rm h}$ $\simeq$ $13~h^{-1}$ M$_{\odot}$) and galaxy clusters (log$_{10}$ $M_{\rm h}$ $\simeq$ $[14,15]~h^{-1}$ M$_{\odot}$). 
The maximum $V_{\rm max}$ values shown in each panel are the maximum ones corresponding to each chosen $M_{\rm h}$ interval. For each case, we have divided the log $V_{\rm max}$ interval in three bins of equal size for each $M_{\rm h}$ mass range and simulation. As can be seen from the resulting values shown in the plot legends, there are no subhalo $V_{\rm max}$ intervals covered by more than one simulation and so it is not possible to use a fixed set of bins across all panels.
For ShinUchuu and Uchuu, we can see that the dependency of $c_{\rm V}$ on $V_{\rm max}$ increases as the mass of the host halo increases and, more importantly, as the range of probed $V_{\rm max}$ values gets broader. Indeed, we find no significant dependence of $c_{\rm V}$ on $V_{\rm max}$ for subhaloes in Phi-4096 (except perhaps in the intermediate host halo mass bin considered), the reason being the comparatively small $V_{\rm max}$ range covered in this case. 
Overall, the behaviour of the data in the different panels of Fig.~\ref{fig:Cv_xbin_halomass_vmax} can be well understood by the implicit dependence of $c_{\rm V}$ on $V_{\rm max}$, shown e.g. in Fig.~\ref{fig:cv-Vmax-z0}: at the lowest $V_{\rm max}$ values, i.e. those probed by Phi-4096, such dependence is weak and thus similar $c_{\rm V}$ values are expected in all cases. As the host halo mass increases and we start probing larger $V_{\rm max}$ subhalo values, the $c_{\rm V}$ -- $V_{\rm max}$ dependence becomes stronger, power law-like. This, coupled with the broader $V_{\rm max}$ bins used for ShinUchuu and Uchuu, translates into a significant and appreciable change of  $c_{\rm V}$ values in the middle and right panels of Fig.~\ref{fig:Cv_xbin_halomass_vmax}, especially for the case of the largest host halo masses and/or the broadest $V_{\rm max}$ ranges considered.
%
Note, also, that the results shown in Fig.~\ref{fig:Cv_xbin_halomass} are in good agreement with those found in Fig.~\ref{fig:Cv_xbin_halomass_vmax}, this way reaffirming our conclusion that subhaloes --at a given $V_{\rm max}$-- within more massive host haloes possess, on average, higher concentrations than those residing in less massive ones.

\section{Evolution of subhalo properties with cosmic time}
\label{sec:redshift}

\subsection{Abundances and radial distribution}
A fundamental issue to understand in detail the process of structure formation in our Universe is the evolution of abundances and concentrations of subhaloes over cosmic time. This is a purely gravitational problem where accretion, mergers, dynamical friction and tidal stripping take place and, as such, is ideally suited for N-body simulations. Thanks to the large volume, high mass resolution and superb statistics of the simulations used in this work, we can analyze the evolution of subhalo properties with redshift in great detail for different host halo masses. 
We expect haloes of a given mass to contain more subhaloes at earlier times since subhaloes in present-day haloes fell into their parent systems more recently. Different works have found this trend using cosmological simulations~\citep{Gao:2004,Gao:2011, Ishiyama2013}.

\begin{figure*}
\centering
\includegraphics[width=\textwidth]{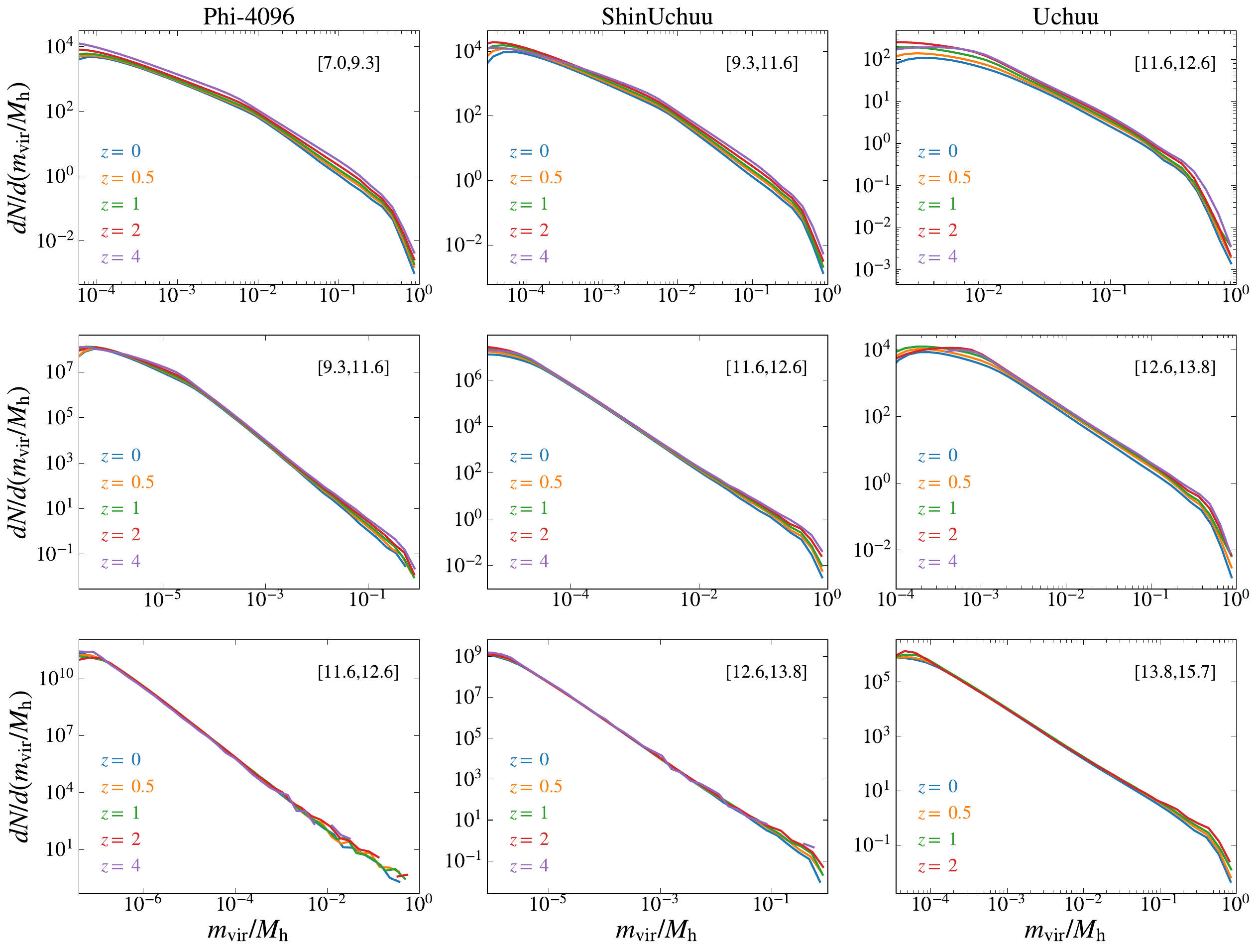}
\caption{
SHMF of the Phi-4096, ShinUchuu and Uchuu simulations,
from left to right, and for different host halo masses, growing from top to bottom. In each panel different redshifts, from 0 to 4, are shown in different colours according to the legend. 
The $x$ axis in each panel represents the ratio between the subhalo mass and its host mass. 
}
\label{fig:shmfzale}
\end{figure*}

\begin{figure*}
\centering
\includegraphics[width=\textwidth]{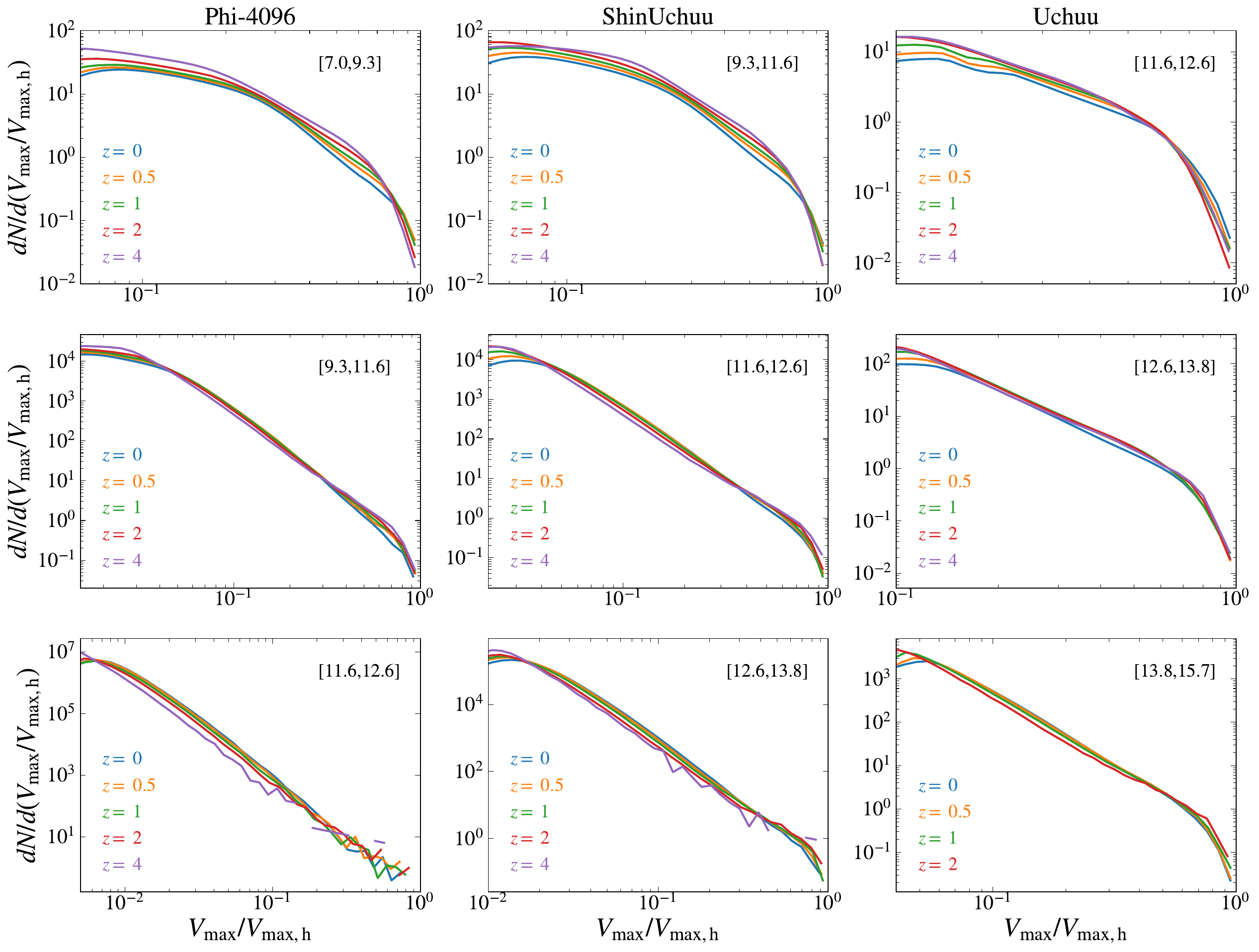}
\caption{
SHVF of the Phi-4096, ShinUchuu and Uchuu simulations, from left to right, and for different host halo masses, growing from top to bottom. In each panel different redshifts, from 0 to 4, are shown in different colours according to the legend. 
The $x$ axis in each panel represents the ratio between the subhalo $V_\mathrm{max}$ and its host $V_\mathrm{max}$.}
\label{fig:shvfzale}
\end{figure*}

The redshift dependence of the SHMF and SHVF obtained from our set of  simulations are shown in Fig.~\ref{fig:shmfzale} and Fig.~\ref{fig:shvfzale}, respectively. 
For each simulation, we have binned the data in three different host mass intervals and repeated the bins used in Section~\ref{sec:shprop}. 
We found that the SHMF evolves weakly with redshift in all three simulations. 

As we did for $z=0$ in Fig.~\ref{fig:shmfmwalegao}, we also checked that, at each redshift shown in Figs.~\ref{fig:shmfzale} and \ref{fig:shvfzale}, more massive hosts possess a larger number of subhaloes.
We note that for the most massive host haloes considered we obtain the expected power-law behaviour of the SHMF over nearly seven orders of magnitude in $m_{\rm vir}/M_\mathrm{h}$ in the case of Phi-4096, six for ShinUchuu and five for Uchuu (bottom panels of Fig.~\ref{fig:shmfzale}). In the case of the SHVF, these subhalo-to-host mass ratios roughly correspond to 2.5, 2 and 1.5 orders of magnitude in subhalo-to-host $V_{\rm max}$ ratio, respectively (bottom panels of Fig.~\ref{fig:shvfzale}). In all cases, the indices of the power laws (see Eq.~\ref{eq:rp16m}) agree well with expectations, e.g. the obtained values range between $1.8-1.9$ for the SHMF. These power-law indices seem to be independent on redshift.

\begin{figure*}
\centering
\includegraphics[width=\textwidth]{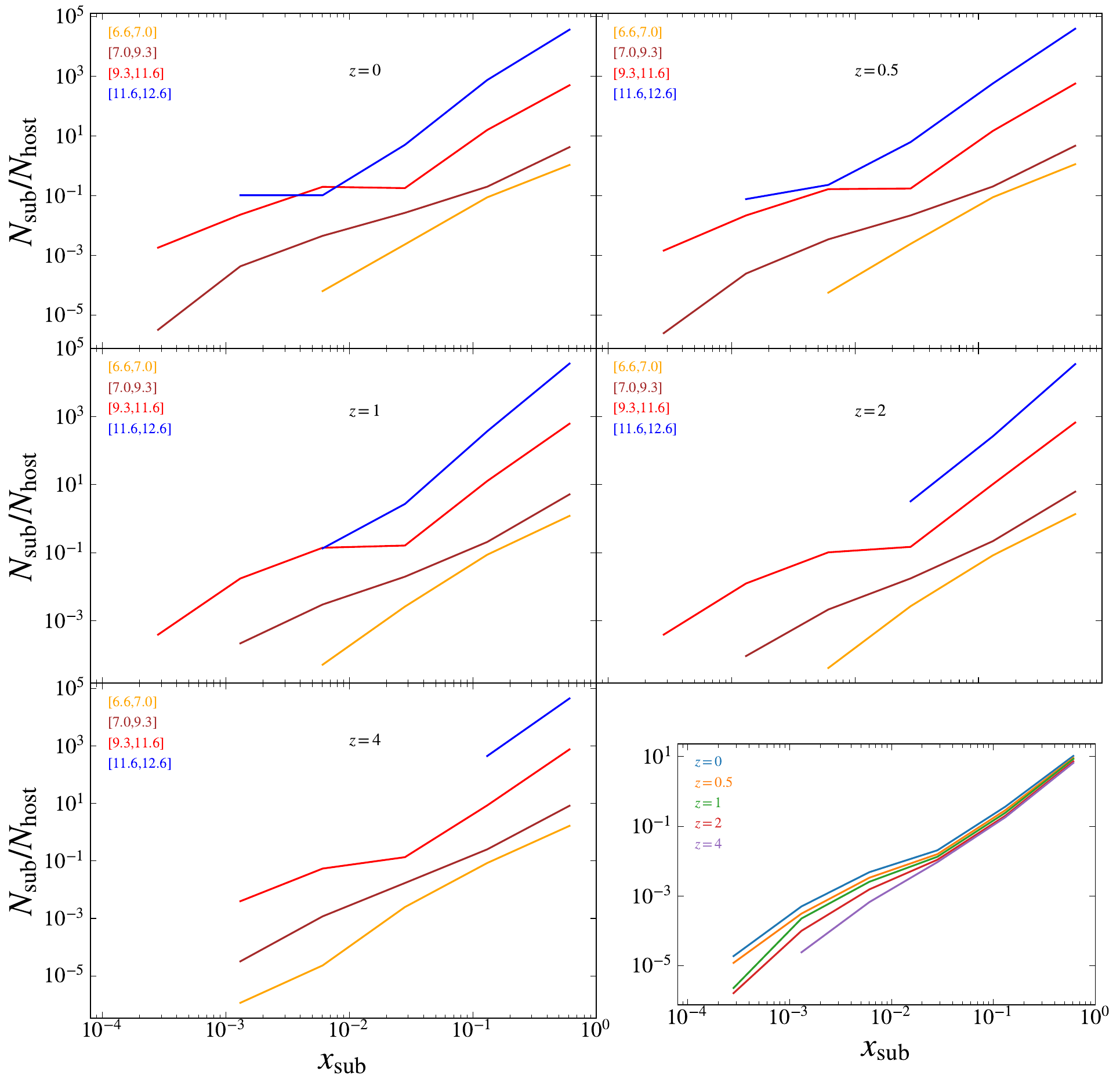}
\caption{SRD of the Phi-4096 simulation for different host halo masses (shown with different colours according to the legends) and different redshifts. From top to bottom and left to right, from  $z=0$ to $z=4$. The last panel on the bottom right shows the resulting SRD without distinction on host halo mass.}
\label{fig:srdphiale}
\end{figure*}

\begin{figure*}
\centering
\includegraphics[width=\textwidth]{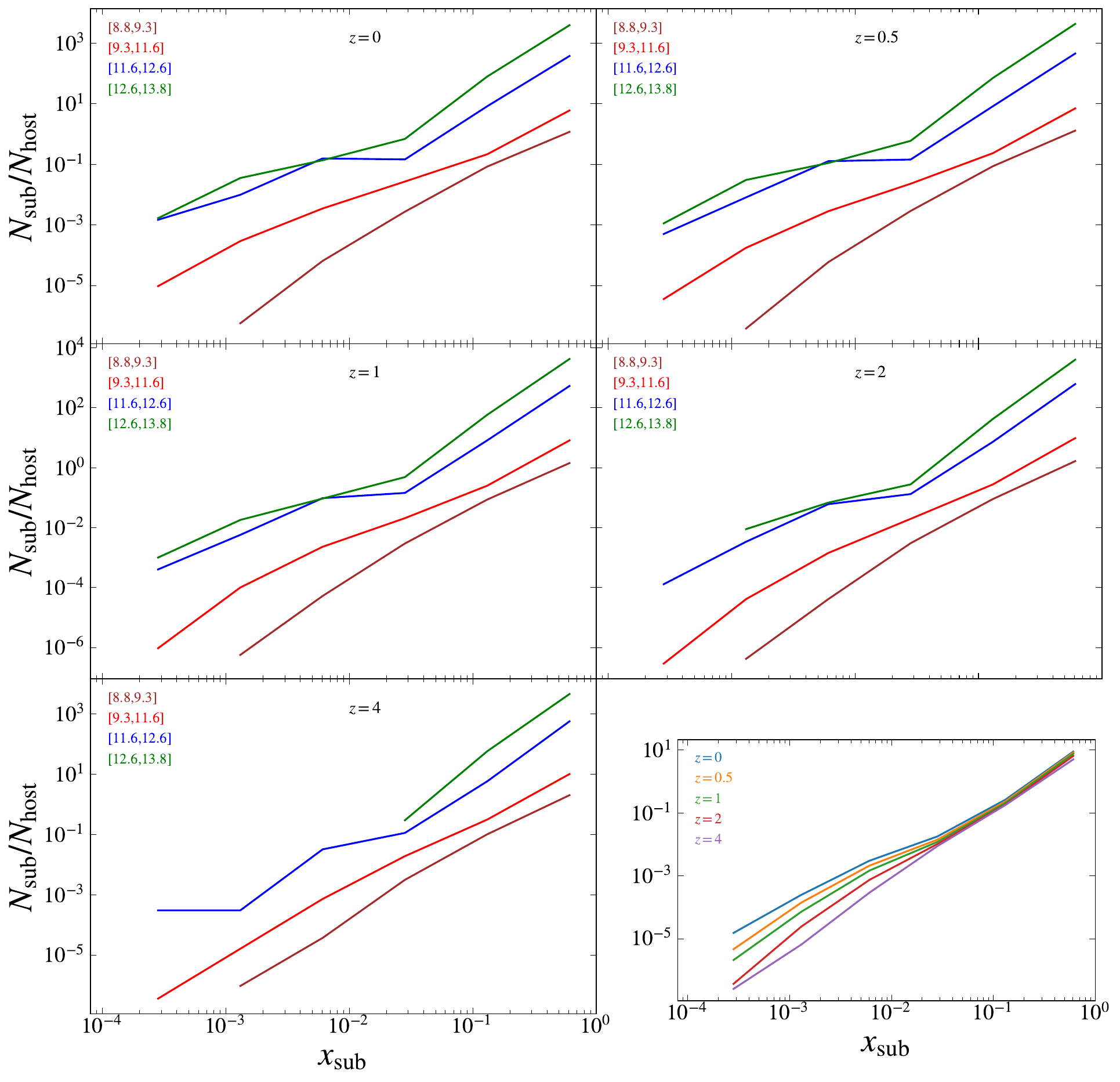}
\caption{SRD of the ShinUchuu simulation for different host halo masses (shown with different colours according to the legends) and different redshifts. From top to bottom and left to right, from  $z=0$ to $z=4$. The last panel on the bottom right shows the resulting SRD without distinction on host halo mass. }
\label{fig:srdsucale}
\end{figure*}

\begin{figure*}
\centering
\includegraphics[width=\textwidth]{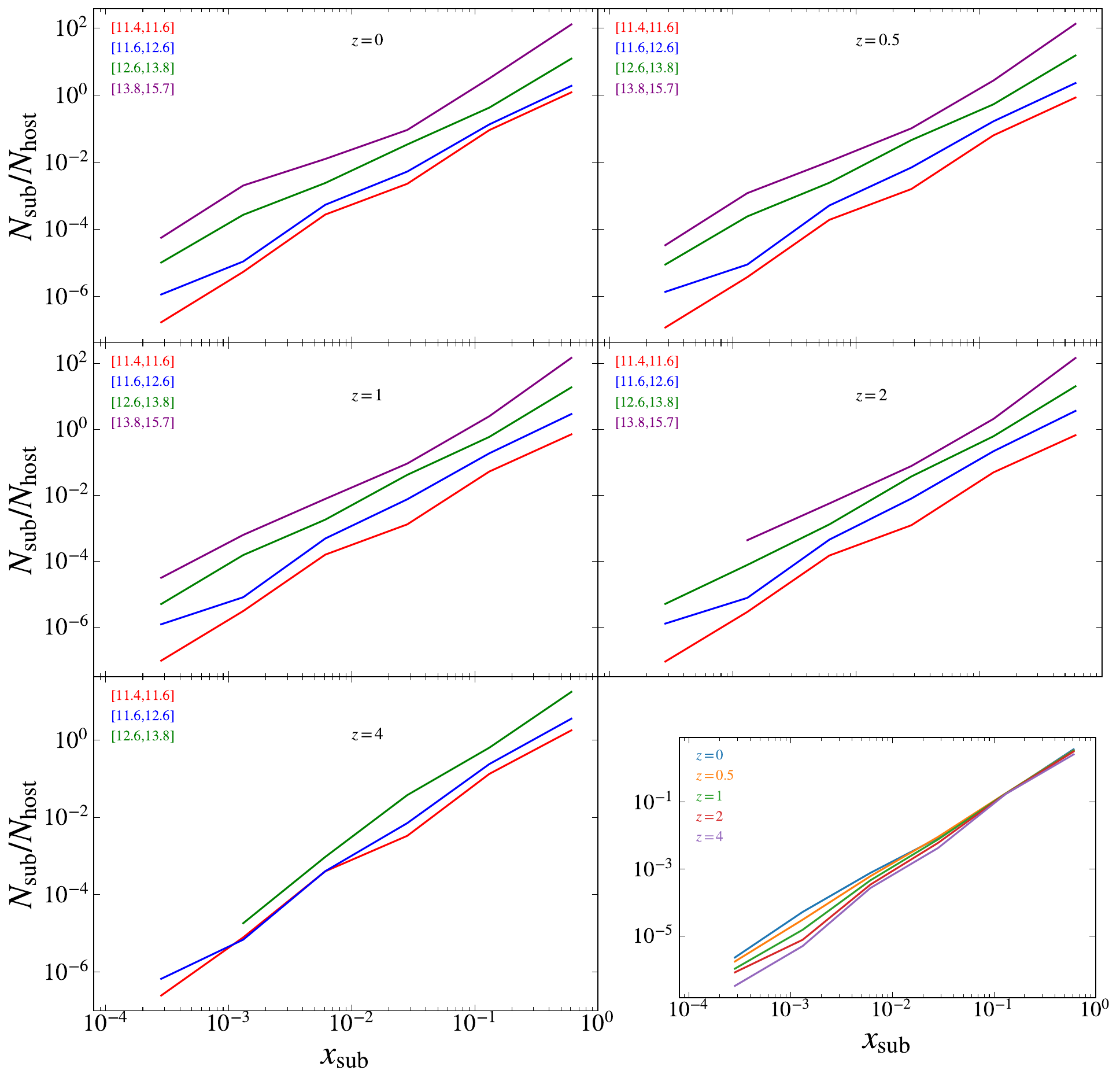}
\caption{SRD of the Uchuu simulation for different host halo masses (shown with different colours according to the legends) and different redshifts. From top to bottom and left to right, from  $z=0$ to $z=4$. The last panel on the bottom right shows the resulting SRD without distinction on host halo mass. }
\label{fig:srduchuuale}
\end{figure*}

We also studied the evolution of the SRD with time. The SRD has been built using the same four host halo bins used in Section~\ref{sub-sec:abundances} above, when calculating the SHMF and SHVF at $z=0$, with six logarithmically equispaced radial bins 
between $x=10^{-4}$ and 1. We show the obtained SRDs for different redshifts in Figs.~\ref{fig:srdphiale}, \ref{fig:srdsucale} and \ref{fig:srduchuuale} for Phi-4096, ShinUchuu and Uchuu, respectively.
In addition to finding, for each SRD, the same overall behaviour already shown in Fig.~\ref{fig:srdz0} for $z=0$, 
the panels of Figs.~\ref{fig:srdphiale} to \ref{fig:srduchuuale} just reflect the well-known hierarchical character of DM halo build-up in $\Lambda$CDM in different ways, with smaller haloes and their subhaloes forming first, and then more massive haloes and their subhaloes only existing at more recent times. In general, we find a larger number of subhaloes in larger hosts, as expected.
In each of these figures, the lower right panel shows the time evolution of the SRD built from all subhaloes in the simulation, independently of host halo mass. Overall, we find more subhaloes at later times, especially in the innermost regions of the hosts. 

\subsection{Concentrations}
\label{subsec:concentration-z}
\begin{figure*}
\centering
\includegraphics[width=0.8\textwidth]{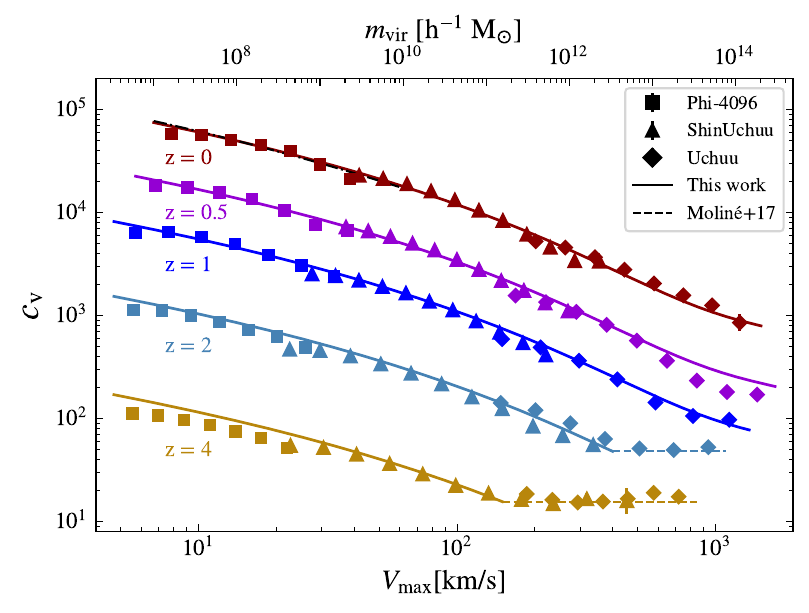}
\caption{Median subhalo concentration $c_{\rm V}$, and the standard error of the median as a function of $V_{\rm max}$ (virial masses along the x-axis on the top) as found in the Phi-4096, ShinUchuu  and the Uchuu simulations. Results for subhaloes depicted at  different redshifts are shown. From top to bottom: $z \,=\, 0$ (red symbols), $z \,=\,0.5$ (violet), $z \,=\,1$ (blue), $z \,=\,2$ (light blue) and $z \,=\,4$ (gold). Solid lines correspond to our fits (coloured lines) given by Eq.~\ref{eq:cv-z-fit} for each of the five considered redshifts. For redshifts $2$ and $4$ and at the highest $V_{\rm max}$ values, we use a constant $c_{\rm V}$ value in each case (thin dashed lines); see text for details.
As a reference, we also show at $z\,=\,0$ the \citet{Moline17} parametrization (dash-dot line), derived only from subhalo data in Milky-Way-like hosts.}
\label{fig:cv_Vmax-z}
\end{figure*}

The redshift dependence of subhalo concentrations in our simulations has also been studied in detail. In Fig.~\ref{fig:cv_Vmax-z} we present the $V_{\rm max}$-$c_{\rm V}$ relation, the latter obtained as in Eq.~\ref{eq:cv-def} at five different redshifts, $z \,=\, 0, \,0.5, \,1, \,2$ and $4$. For each redshift, we show median $c_{\rm V}$ values and corresponding errors for Phi-4096, ShinUchuu and Uchuu.  
The same resolution criteria implemented for the cuts in Section~\ref{sub-sec:concentrations} for $z = 0$ have been applied to the $V_{\rm max}$ values of both haloes and subhaloes in each simulation. As a result, we show the concentration of subhaloes with $V_{\rm max}$ above 7 km s$^{-1}$ for $z = 0$, 6 km s$^{-1}$ for $z = 0.5$, and 5 km s$^{-1}$ for $z = 1,\,2$ and $4$, which are the cuts applied on the Phi-4096 simulation.
\footnote{In the case of ShinUchuu we use haloes and subhaloes with  $V_{\rm max} \geq 38$ km s$^{-1}$ at $z = 0$, $V_{\rm max} \geq 34$ km s$^{-1}$ at $z = 0.5$, $V_{\rm max} \geq 25$ km s$^{-1}$ at $z = 1$ and $V_{\rm max} \geq 20$ km s$^{-1}$ at $z = 2$ and $4$. For Uchuu, we consider subhaloes with $V_{\rm max} \geq 180$ km s$^{-1}$ for $z = 0$, $V_{\rm max} \geq 150$ km s$^{-1}$ for $z = 0.5$,  $V_{\rm max} \geq 130$ km s$^{-1}$ for $z = 1, \,2$ and $V_{\rm max} \geq 170$ km s$^{-1}$ for $z = 4$. In this case, the subhaloes reside in host haloes with $V_{\rm max,h} \geq 270$ km s$^{-1}$ at $z = 0$, $V_{\rm max.h} \geq 160$ km s$^{-1}$ at $z = 0.5,\,1, \,2$ and $V_{\rm max,h} \geq 170$ km s$^{-1}$ at $z = 4$.} We see the expected result that the median subhalo concentration declines with increasing mass and redshift. The shape of the $V_\mathrm{max}$-$c_{\rm V}$ median relation also evolves with redshift. Indeed, for $z > 1$, we find this relation to flatten and to remain practically constant (or even slightly increasing) at the highest masses. Similar results were found for haloes where an upturn or flattening in their concentrations was obtained at high mass and redshift~\citep{Zhao:2003,Gao:2008,Zhao:2009,Bolshoi,Munoz-Cuartas:2011,Prada:2012,Diemer:2015,Correa:2015,BolshoiP16,RP:2016,Child:2018,Diemer2019,Ishiyama2021}. The reason for this upturn is uncertain. A possible explanation is related to the precise statistics of the highest density peaks~\citep{Bolshoi,Child:2018,Ishiyama2021}. Non-equilibrium effects have also been proposed, since haloes in the upturn are those with the largest masses at any given moment and, thus, are
known to grow very fast (see, e.g. \citealt{Ludlow:2014}). Yet, analyses from simulations reveal that out-of-equilibrium effects may not provide a convincing explanation, since selecting relaxed haloes only increases the magnitude of the upturn~\citep{Prada:2012}. Other works see no upturn using relaxed haloes and focus on discussing differences in methodology when deriving concentrations and corresponding relations with mass (or equivalently, $V_{\rm max}$) and redshift~\citep{Ludlow:2012,Ludlow:2014,Hellwing:2015upa,Angel:2016}. Here, we prefer not to perform a detailed analysis of subhaloes in this potential upturn or plateau at high masses and redshifts, and postpone its discussion to future work.

The data and results shown in Fig~\ref{fig:cv_Vmax-z} for different redshifts allow us to obtain a parametrization for $c_{\rm V}$ as a function of $V_{\rm max}$ and $z$ for subhaloes:
\begin{eqnarray}
c_{\rm V}(V_{\rm max},z) & = & c_{0} \, \left[1 + \sum_{i=1}^{3} \,  a_{i} \, \left[\log_{10}\left(\frac{V_{\rm max}}{{\rm km \, s^{-1}}}\right)\right]^{i}\right] \times  \nonumber \\ 
& & \left[\left(1 +  z \, \right)^b  \left(1 + d z \, \right)\right] ~,
\label{eq:cv-z-fit}
\end{eqnarray}
where $c_{0} = 1.75 \times 10^5$ and $a_{i}=\left\{\color{blue!70!black}{-0.90368, \, 0.2749, \, -0.028} \right\}$, $b = -5.52$, $d = 3.2$. This fit works well --with an error smaller than $\sim$5 per cent-- in the subhalo $V_{\rm max}$ range $7$~km s$^{-1}$ $\lesssim V_{\rm max} \lesssim 1500$~km s$^{-1}$ for $z=0$, ~$6$~km s$^{-1}$ $\lesssim V_{\rm max} \lesssim 1700$~km s$^{-1}$ for $z=0.5$, ~$5$~km s$^{-1}$ $\lesssim V_{\rm max} \lesssim 1350$~km s$^{-1}$ for $z=1$, ~$5$~km s$^{-1}$ $\lesssim V_{\rm max} \lesssim 400$~km s$^{-1}$ for $z=2$ and ~$5$~km s$^{-1}$ $\lesssim V_{\rm max} \lesssim 150$~km s$^{-1}$ for $z=4$. 
 Our fits for different redshifts as given by Eq.~\ref{eq:cv-z-fit} are shown in Fig~\ref{fig:cv_Vmax-z} as solid coloured lines. The \cite{Moline17} parametrization is also shown for comparison at $z=0$. Here, and for the reasons already explained in Section~\ref{sub-sec:concentrations} (Fig.~\ref{fig:cv-Vmax-z0}), we obtain a slope similar to that in~\cite{Moline17}, yet with differences at high maximum circular velocities ($V_{\rm max} > 150$ km s$^{-1}$). 

As mentioned above, in this work we do not study in detail the properties of those subhaloes lying in the $c_{\rm V}$ plateau at high redshifts and masses. Thus, we do not include their corresponding median concentration values in the data set that we used to obtain our best-fitting parameters. Instead, we provide here constant $c_{\rm V}$ values for the plateaus at redshifts $2$ and $4$, depicted as a thin dashed horizontal line in Fig~\ref{fig:cv_Vmax-z}. Both are the result of simply evaluating our parametrization of Eq.~\ref{eq:cv-z-fit} at the maximum $V_{\rm max}$ considered for each redshift: $c_{\rm V}(400,2)=47.98$ and $c_{\rm V}(150,4)=15.56$, respectively. 


\section{Summary and conclusions}
\label{sec:conc}

In this work, we have studied in detail the subhalo population using a combination of state-of-the-art N-body cosmological simulations, namely the large-scale Uchuu simulation suite and the Phi-4096 extremely-high resolution simulation. 
The superb subhalo statistics, together with both the large volume and high-mass resolution described in Section~\ref{sec:simu}, allowed us to characterize both the abundance and structural properties of subhaloes over various decades of the subhalo-to-host-halo mass ratio, for the first time consistently for host haloes of very different masses (seven, six and four orders of magnitude, for example, for MW-size hosts, galaxy groups and galaxy clusters, respectively). First, in Section~\ref{sec:shprop} we dissected the abundance of subhaloes as well as their distribution within the hosts and concentrations as a function of mass, subhalo maximum circular velocity and distance to the host halo centre at $z=0$. We also investigated the dependency of these subhalo properties on host halo mass. Then, in Section~\ref{sec:redshift} we analyzed the evolution of all these dependencies with cosmic time, reaching $z=4$ with yet great statistics.

In particular, in the context of subhalo abundance, we built the differential subhalo mass function in the range between $10^{4} - 10^{15.2}~h^{-1}$ M$_{\odot}$ and for host halo masses $10^{6.6} - 10^{15.7}~h^{-1}$ M$_{\odot}$. We also derived the subhalo velocity function between $1-1874$ km s$^{-1}$, and for host haloes with maximum circular velocities between $4.5-2582$ km s$^{-1}$. The radial distribution of subhaloes within their hosts was also obtained using subhaloes located at distances as deep as just $\sim 0.1$ per cent the virial radius of the host.
Subhalo concentrations, $c_{\rm V}$, were calculated in our work independently of any pre-defined density profile and built only in terms of the more physical $V_{\rm max}$ and $R_{\rm max}$ parameters, this way avoiding any potential issues from having tidally-stripped profiles for subhaloes. A more rigorous data selection was also applied in this case in order to avoid resolution issues at the smallest considered scales. Even with our highly-demanding quality cuts, we were able to obtain concentration results for subhaloes with masses in the range $10^6 - 10^{14}~h^{-1}$ M$_{\odot}$ residing in host haloes with masses $10^7 - 10^{15}~h^{-1}$ M$_{\odot}$. All together, our careful simulation analysis work made it possible to extend, by several orders of magnitude in mass, both at the high- and low-mass ends, previous results on subhalo abundances and concentrations. The main results of our work can be summarized as follows.

\begin{enumerate}
\item The slopes of both our SHMF and SHVF are in agreement with previous results and expectations, and show no dependence on redshift. This SHMF depends weakly on host halo mass (see Fig.~\ref{fig:shmfzeroale}). More precisely, at $z=0$ we find up to a factor $\sim 2-3$ more subhaloes in galaxy-cluster-size haloes compared to those in dwarf-galaxy-size hosts.
All three simulations are in good agreement with each other, 
our SHMF best-fitting parameters agreeing well with the simulation data at better than $\sim20$ per cent for Phi-4096 and ShinUchuu, and $\sim40$ per cent for Uchuu; for the SHVF, the differences are less than $\sim10$ per cent for Phi-4096 and ShinUchuu,
and up to $\sim50$ per cent for Uchuu (see Fig.~\ref{fig:shmfmwale}). 
We also find more subhaloes at higher redshifts for the same host halo mass (see Figs.~\ref{fig:shmfzale} and~\ref{fig:shvfzale}).


\item In addition to beautifully showing in detail the well-known hierarchical assembly of structures with time as it happens within the standard cosmological scenario, our study of the SRD with redshift, illustrated by Figs.~\ref{fig:srdphiale}, \ref{fig:srdsucale} and \ref{fig:srduchuuale}, confirms the existence of a larger number of subhaloes in more massive hosts, as expected, and shows no significant variation of this number with time. Also, as time evolves, subhaloes fall deeper into the inner parts of their hosts. 
We stress that, in this work, we were able to follow the evolution of the SRD with unprecedented detail and consistently over more than seven decades in subhalo mass, for subhaloes located as deep as just 0.1 per cent of the virial radius of their hosts.

\item  Qualitatively, and as already presented in previous works, we found the subhalo concentration a) to slowly decrease with increasing subhalo mass and b) to significantly increase towards the host halo centre for subhaloes of the same mass (Fig.~\ref{fig:cv-Vmax-z0}). For the first time, we consistently characterize the $c_\mathrm{v}-V_{\rm max}$ relation for subhaloes in the wide range $7-1500$ km s$^{-1}$, and provide a new parametrization that includes the dependence on distance to host halo centre. This parametrization represents a significant improvement with respect to the one presented in~\cite{Moline17}, that was based on subhalo data only from Milky-Way-like hosts. 
In particular, we found the innermost, less massive subhaloes in our simulations to exhibit $c_\mathrm{v}$ values up to a factor $\sim 3$ higher than those located in the outermost regions of their hosts, being this difference of just a factor $\sim 1.5$ for the most massive subhaloes. 

\item Interestingly, we found subhaloes of the same mass to be significantly more concentrated when they reside inside more massive hosts (Fig.~\ref{fig:Cv_xbin_halomass}). We found no explicit mention to this effect in the previous literature.



\item The redshift dependence of subhalo concentrations in our simulations showed the expected result that the median of subhalo concentrations declines with increasing both the subhalo $V_{\rm max}$ and redshift. Yet, at the highest considered masses and for redshifts above one, we found that the concentration flattens and then remains practically constant -- or even increases slightly with subhalo mass (Fig.~\ref{fig:cv_Vmax-z}). In our work, we provided the first accurate fit (Eq.~\ref{eq:cv-z-fit}) for the time evolution of subhalo concentrations for a large range of subhalo $V_{\rm max}$ values and valid at least up to $z=4$.

\end{enumerate}

The results in this work offer an unprecedented, detailed characterization of the DM subhalo population. Improving our knowledge about the latter is of prime importance since subhaloes represent important probes of the mass accretion history and dynamics of host haloes and thus, ultimately, of the underlying cosmological model. We also expect our results to be particularly useful to shed light on the actual role of subhaloes in dark matter searches. Indeed, some of the results in this work can be critical in this regard, as they represent a qualitative leap with respect to previous numerical efforts in this same direction.

\section*{Acknowledgements}

The work of AM, MASC and AAS was supported by the Spanish Agencia Estatal de Investigaci\'on through the grants PGC2018-095161-B-I00 and IFT Centro de Excelencia Severo Ochoa SEV-2016-0597, the {\it Atracci\'on de Talento} contracts no. 2016-T1/TIC-1542 and 2020-5A/TIC-19725 granted by the Comunidad de Madrid in Spain, and the MultiDark Consolider Network FPA2017-90566-REDC. The work of AM was also supported by "Comunidad de Madrid S2018/NMT-4291  TEC2SPACE-CM". The work of AAS was also supported by the Spanish Ministry of Science and Innovation through the grant FPI-UAM 2018. FP thanks the support of the Spanish Ministry of Science and Innovation funding grant PGC2018- 101931-B-I00.
TI was supported by MEXT as "Program for Promoting Researches on the Supercomputer Fugaku'' (JPMXP1020200109),  MEXT/JSPS KAKENHI Grant Number JP19KK0344, JP20H05245, JP21H01122 and IAAR Research Support Program, Chiba University, Japan. SAC acknowledges funding from
 Consejo Nacional de Investigaciones Cient\'{\i}ficas y T\'ecnicas (CONICET, PIP-0387), 
Agencia Nacional de Promoci\'on de la Investigaci\'on, el Desarrollo Tecnol\'ogico y la Innovaci\'on (Agencia I+D+i, PICT-2018-3743), and  Universidad Nacional de La Plata (G11-150), Argentina. EJ acknowledges financial support from CNRS. TO is supported by JSPS KAKENHI Grant Number 20K22360 and 21H05449.

This work used the skun6@IAA facility (\url{http://www.skiesanduniverses.org}) managed by the Instituto de Astrof\'{i}sica de Andaluc\'{i}a (CSIC). The equipment was funded by the Spanish Ministry of Science EU-FEDER infrastructure grants EQC2018-004366-P and EQC2019-006089-P. The Uchuu, Shin-Uchuu, and Phi-4096 simulations were carried out on Aterui II supercomputer at Center for Computational Astrophysics, CfCA, of National Astronomical Observatory of Japan. 
The numerical analysis were partially carried out on XC40 at the Yukawa Institute Computer Facility in Kyoto University.

\section*{Data availability}


The data underlying this article were accessed from the Skies \& Universes site: \url{http://www.skiesanduniverses.org}. The derived data generated in this research will be shared on reasonable request to the corresponding author.
\bibliography{biblio}
\bibliographystyle{astron}

\appendix
\section{Selection cuts on subhalo maximum circular velocities}
\label{sec:appendix}
In this Appendix, we detail the criteria we used to define and apply additional and specific cuts on the subhalo maximum circular velocity data in order to avoid resolution issues that may impact the determination of the subhalo concentrations. 

\begin{figure}
\includegraphics[width=0.47\textwidth]{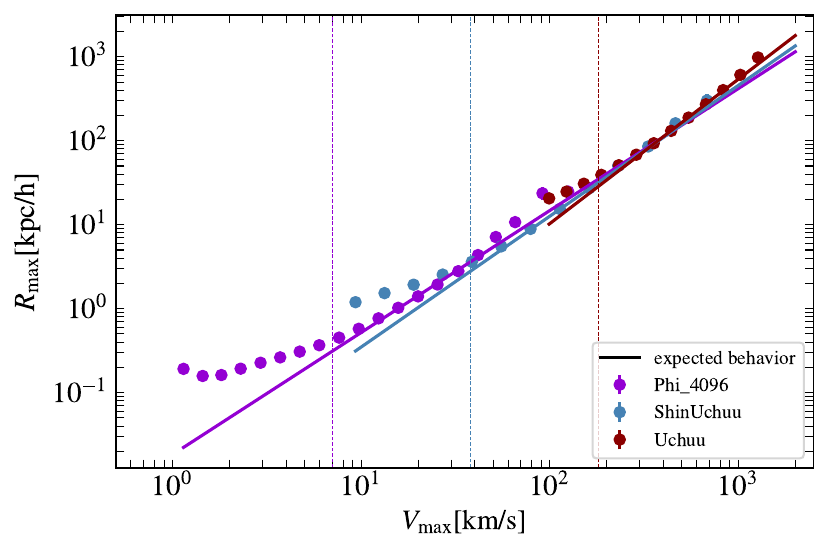}
\caption{Median  $V_{\rm max}-R_{\rm max}$ values in the Phi-4096 (violet), ShinUchuu (light blue) and Uchuu (red) simulations. Solid lines correspond to the expected linear behaviour for each simulation. Vertical dashed lines indicate the cuts in the data applied in this work for each simulation, i.e. only subhaloes with $V_{\rm max} \geq 7$ km s$^{-1}$ are included in our analyses for Phi-4096, {$V_{\rm max} \geq 38$ km s$^{-1}$ for ShinUchuu, and $V_{\rm max} \geq 180$ km s$^{-1}$ for Uchuu. Below the mentioned cut value the $V_{\rm max} - R_{\rm max}$ relation is no longer linear}. }
\label{fig:appendix}
\end{figure}

Fig.~\ref{fig:appendix} shows the $V_{\rm max} - R_{\rm max}$ relation for all subhaloes found at redshift $z = 0$ for each simulation used in this work. The expected behaviour of this relation is almost linear as was studied in different works (see e.g.~\citealt{Xu:2015,Springel08}).  We show both the medians and the expected linear $V_{\rm max} - R_{\rm max}$ behaviour found for each simulation after fitting the data to a linear function.
In this figure we can see that the Vmax values at which the $V_{\rm max} - R_{\rm max}$ relation is no longer linear are those below the ones given in Tab.~\ref{tab:Cv-minmax}.  After applying the cuts, from Fig.~\ref{fig:appendix} we see that the resulting data is in good agreement across all the simulations of our suite in the interval where their  $V_{\rm max}$ values overlap.
 
As a double check, we calculated the concentrations of the subhaloes with $V_{\rm max}$ values below the cut chosen for each simulation. We found that their behavior is non-physical, with the concentrations decreasing as Vmax decreases from the values just below our cuts.
\end{document}